\documentclass[useAMS,usenatbib]{mn2e}
\usepackage{amsmath,fleqn,graphicx,amssymb}
\usepackage{multirow}
\renewcommand{\[}{\begin{equation}}
\renewcommand{\]}{\end{equation}}
\def\p{\partial}\def\i{\rm i}

\let\boldgrk=\gkvecten
\let\boldgrksc=\gkvecseven

\def\gkthing#1{{\mathchoice%
	{\hbox{{\boldgrk\char#1}}}
	{\hbox{{\boldgrk\char#1}}}
	{\hbox{{\boldgrksc\char#1}}}
	{\hbox{{\boldgrksc\char#1}}}}}

\def\vtheta{\gkthing{18}}

{\newif\ifnotend
\notendtrue
\def\veclist{ABCDEFGHIJKLMNOPQRSTUVWXYZabcdefghijklmnopqrstuvwxyz.}
\def\top#1#2.{#1}
\def\tail#1#2.{#2.}
\loop\expandafter\xdef\csname v\expandafter\top\veclist\endcsname%
{{\noexpand\bf\expandafter\top\veclist}}
\edef\veclist{\expandafter\tail\veclist}
\if\veclist.\notendfalse\fi\ifnotend\repeat}

\def\df{\textsc{df}}

\def\bolOm{\mbox{\boldmath$\Omega$}}
\def\vOmega{\bolOm}
\def\vDelta{\mbox{\boldmath$\Delta$}}

\def\Gyr{\,\mathrm{Gyr}}
\def\Myr{\,\mathrm{Myr}}
\def\kpc{\,\mathrm{kpc}}

\def\cut{\mathrm{cut}}

\def\kms{\,\mathrm{km\,s}^{-1}}

\def\bolJ{\mbox{\boldmath{$J$}}}

\def\vlos{{v_\parallel}}

\def\rd{\mathrm{d}}\def\e{\mathrm{e}}
\def\Rc{R_\mathrm{c}}\def\Rd{R_\mathrm{d}}
\def\fracj#1#2{{\textstyle{#1\over#2}}}

\def\codename{\textsc{tm}}
\def\rs{r_{\rm s}}

\def\tolJ{\textsc{tol}_J}
\def\rms{\textsc{rms}}
\def\vthetaT{\vtheta^{\rm T}}\def\vJT{\vJ^{\rm T}}
\def\thetaT{\theta^{\rm T}}\def\JT{J^{\rm T}}\def\PhiT{\Phi^{\rm T}}
\def\vxT{\vx^{\rm T}}\def\vvT{\vv^{\rm T}}
\def\rT{r^{\rm T}}\def\varthetaT{\vartheta^{\rm T}}\def\pT{p^{\rm T}}
\title[Torus mapping code]
{Torus mapper: a code for dynamical models of galaxies}

\author[James Binney \& P.~J.~McMillan]{
  James Binney$^1$\thanks{E-mail: binney@thphys.ox.ac.uk} and Paul~J.~McMillan$^{1,2}$\\  
  $^1$Rudolf Peierls Centre for Theoretical Physics, 1 Keble Road,
  Oxford, OX1 3NP, UK \\
  $^2$Lund Observatory, Lund University, Department of Astronomy and Theoretical Physics, Box 43, SE-22100, Lund, Sweden \\
}

\begin{document}
\maketitle

\begin{abstract}
  We present a freely downloadable software package for modelling the dynamics of
  galaxies, which we call the Torus Mapper (\codename). 
  The package is based around `torus
  mapping', which is a non-perturbative technique for creating orbital tori
  for specified values of the action integrals. Given an orbital torus and
  a star's position at a
  reference time, one can compute its position at any other time, no matter
  how remote. One can
  also compute the velocities with which the star will pass through any given point
  and the contribution it will make to the time-averaged density there. A
  system of angle-action coordinates for the given potential can be created by 
  foliating phase space with orbital tori. Such a foliation is facilitated by
  the ability of \codename\ to create tori by interpolating on a grid of
  tori.

  We summarise the advantages of using \codename\ rather than a standard
  time-stepper to create orbits, and give segments of code that
  illustrate applications of \codename\ in several contexts, including
  setting up initial conditions for an N-body simulation.
  We examine the precision of the orbital tori created by \codename\ and
  the behaviour of the code when orbits become trapped by a resonance.
\end{abstract}


\section{Introduction} \label{sec:intro}

Dynamical models are playing an increasingly important role in the
interpretation of observations. Moreover, the volume and quality of available
observational data for both external galaxies and our own Galaxy are growing
explosively through the increase in the scale and sensitivity of
large-format CCD detectors and integral field spectrographs, and, in the case
of our own Galaxy, the taking of astrometric data by the Gaia satellite.

N-body models have played a large role in the development of our
understanding of galaxy formation and dynamics, but they have important
limitations. 1) They are significantly degraded by discreteness noise,
which is particularly acute in the case of our Galaxy because the data are
strongly focused on the immediate vicinity of one star, the Sun. 2) Tailoring
an N-body model to a specific body of observations is difficult given the
complex and ill-understood nature of the connection between the initial
conditions that define an N-body simulation and the equilibrium model that
emerges. It is true that the made-to-measure technique
\citep{SyerTremaine,DehnenM2M,MorgantiGerhard} can be used to refine the fit between a
model's predictions and data, but making such adjustments increases the level
of discreteness noise in the model, so the starting point for adjustments
should be reasonably close to the truth.  3) There is no compact and
transparent representation of an N-body model: the model is specified by $6N$
phase-space coordinates, but if the equations of motion are integrated for a
fraction of a dynamical time, all these numbers change while the model
remains the same. This degeneracy of representation makes comparison of
models hard. It also obscures the physical significance of a given model and
therefore restricts the harvest of scientific understanding that can be drawn
from a given set of observations. 4) N-body models with acceptable levels of
discreteness noise are computationally costly, so it is unlikely to be
feasible to find an N-body model that provides an excellent fit to a
sophisticated set of observational data.

N-body models do have one important advantage, namely, that they can be
end-points of simulations of cosmological clustering. This advantage is,
however, strongly tempered by the fact that it is currently not feasible to
follow the dynamics of baryons in such simulations using only basic physics.
Consequently, resort is made to ``subgrid recipes'' that should reproduce, in
a statistical sense, the impact on simulated scales of physical processes
that take place on smaller scales. It is not even known whether the neglected
physics {\it can} be reproduced by such recipes, still less is it clear if
any of
the recipes currently used is reasonably accurate. Consequently, cosmological
simulations of galaxy formation lack rigour and should be considered
instructive ways of producing N-body galaxy models, rather than definitive
exercises.

Decades ago, when the data for galaxies were orders-of-magnitude more sparse
than they now are, data were widely modelled with the Jeans equations
\citep[e.g.][]{BaconSM,BDI,vdMBD}. Some studies still rely on the Jeans equations, but
the technique has strong limitations. 1) It recovers only the low-order
velocity moments, which do not specify the actual (non-Gaussian) velocity
distributions. 2) In practice Jeans modelling is restricted to systems
with distribution functions (\df s) that depend on only energy $E$ and
angular momentum $L$, whereas we know that real \df s typically depend on
three isolating integrals.

As data for external galaxies grew richer, the dominant technique for
modelling galaxies became that introduced by \cite{Sc79} and extended by
\cite{vdMarel1998}, \cite{Gebhardt2003} and others, in which orbits
in a given gravitational potential are assigned weights in such a way that
the fit of a model to observational data is optimised. This technique is
flexible but suffers from many of the shortcomings of N-body models, in
particular a high degree of discreteness noise and the lack of compact and
transparent model specification. The keys to obtaining such a specification
are (a) to assign physically meaningful labels to orbits, and (b) to
establish the density with which phase space is sampled by the orbit library
so the weights orbits are assigned can be converted into the value taken by the
\df\ on each orbit. These tasks can be accomplished in the context of a
classical Schwarzschild model \citep[e.g.][]{Haefner}, but there are distinct
advantages to abandoning the use of orbits in favour of phase-space tori,
using the technique described in this paper.

\subsection{Orbital tori and action-angle coordinates}

Numerical integration of orbits in spherical and axisymmetric potentials
reveals that they are nearly always quasiperiodic. That is, the Fourier
decomposition of the time series $x_i(t)$ of any coordinate that one obtains
by integrating the equations of motion contains only discrete spectral lines,
and the frequencies of these lines can be represented as integer linear
combinations of three {\it fundamental frequencies} $\Omega_i$. From this
finding it follows that these orbits are {\it integrable} in the sense that along
them three independent functions $I_i(\vx,\vv)$ of the phase-space
coordinates are constant \citep{Arnold}. The Jeans theorem \citep{Jeans1915}
tells us that the \df\ of a steady-state galaxy with the given potential can
be taken to be a function $f(I_1,I_2,I_3)$ of these constants of motion.

Since any function $J(I_1,I_2,I_3)$ inherits from the $I_i$ the property of
being a constant of motion, there is considerable freedom in what functions
of $(\vx,\vv)$ one uses for the arguments of the \df. However, if one requires
that the adopted functions $J_i$ are capable of being complemented by
canonically conjugate coordinates $\theta_i$, so together
$(\vtheta,\vJ)$ form a system of canonical coordinates, then almost all the
freedom in the choice of constants of motion disappears. If one further
specifies that $J_r$ quantifies the amplitude of a star's oscillations in
radius and $J_z$ quantifies the amplitude of a star's oscillations in the
latitudinal direction (along $\vartheta$ in conventional spherical polar
coordinates), then the last vestiges of freedom disappear and one has a
uniquely defined set of canonical coordinates, with the momenta labelling
orbits and the conjugate variables specifying position within the
orbit.\footnote{$J_r$ is sometimes denoted $J_R$ or $J_u$. Similarly, $J_z$
is identical to what in \cite{GDII} is denoted $J_\theta$. These alternative
notations arise through solving the Hamilton-Jacobi equation in different
coordinate systems: unique underlying actions $J_r$, $J_z$ are being
approached through differing limiting forms.} 

It follows trivially from Hamilton's equations of motion, that angle
coordinates increase linearly in time: $\theta_i(t)=\theta_i(0)+\Omega_it$.
Hence stellar dynamics becomes trivial once the mapping
$(\vtheta,\vJ)\rightarrow(\vx,\vv)$ has been constructed.

The prerequisite of angle-action coordinates $(\vtheta,\vJ)$ playing a useful
role in astronomy is algorithms that alow one to compute $\vx(\vtheta,\vJ)$
and $\vv(\vtheta,\vJ)$, and the inverse functions, $\vJ(\vx,\vv)$ etc.  A
fully analytic algorithm of this type is known only for the multi-dimensional
harmonic oscillator and the isochrone
potential \citep{Henon1959,GDII} (which includes  the Kepler potential as a
limiting case). \cite{McGJJB90} showed how these fully
analytic algorithms can be leveraged into an effective numerical scheme for a
general axisymmetric potential by {\it torus mapping}. The idea is that the
analytic formulae provide an explicit representation of a phase-space surface
$\vJ=\hbox{constant}$. This three-dimensional surface in six-dimensional
phase space is topologically a 3-torus \citep{Arnold}, and this 3-torus is
special in the sense that any two-surface within it has vanishing Poincar\'e
invariant $\sum_i\rd q_i\rd p_i$. Consequently, the orbital surface
$\vJ=\hbox{constant}$ is called a {\it null torus}. If such a null torus can
be mapped into the target phase space such that on it the target Hamiltonian
\[\label{eq:defsH}
H(\vx,\vv)=\fracj12v^2+\Phi(\vx)
\]
 is constant, then the image torus becomes an orbital torus of the target
potential $\Phi(\vx)$.

\cite{McGJJB90} mapped {\it toy tori} labelled by $\vJT$ and furnished with
angle coordinates $\vthetaT$, into tori of a target potential labelled by $\vJ$
with canonical transformations that have generating functions of the form
\[\label{eq:defsS}
S(\vJ,\vthetaT)=\vJ\cdot\vthetaT+\sum s_\vn(\vJ)\e^{\i\vn\cdot\vthetaT}.
\]
 They showed that the image tori became excellent approximations to orbital
tori once the parameters $S_\vn$ had been adjusted to yield a small \rms\
variation $\delta H$ in the Hamiltonian over the image torus.  \cite{Ka94}
showed that certain target tori cannot be reached using generating functions
of this form, and for these tori it is necessary to compound such a
transformation with a point transformation.  The primary purpose of this
paper is to introduce a freely downloadable code, \codename, that computes
these canonical transformations for an arbitrary axisymmetric potential, and
provides a wide range of routines for exploiting the resulting tori.

\subsection{Torus-mapping code TM}

Users of \codename\ do not need to engage with the process of computing
canonical transformations. The only code they need to write is that which
computes the potential and its first derivatives in $R$ and $z$ -- thus
precisely the code required to compute an orbit by numerical integration of
the equations of motion. Given this code and values for the three actions
$J_r$, $J_\phi$ and $J_z$, \codename\ provides functions that return (i)
$(\vx,\vv)$ for specified $\vtheta$, (ii) the Jacobian $\p(\vx)/\p(\vtheta)$,
and (iii) $\vtheta$ given $(\vx,\vJ)$. In Section~\ref{sec:torus} we will see
that the last two functions allow \codename\ to perform the role in
Schwarzschild modelling that has previously been played by an integrator of
the Runge-Kutta, Bulirsch-Stoer, or leap-frog type. Displacing an integrator
with \codename\ brings substantial advantages.

\begin{itemize}
\item[1] A torus is specified by $\sim100$ numbers, namely the $S_\vn$, their
derivatives $\p S_\vn/\p\vJ$, and a handful
of other parameters, rather than by some thousands of phase-space coordinates
along a time sequence, so using \codename\ dramatically shrinks the size of an
orbit library at a given resolution. 

\item[2] It is far easier to construct an orbit library that provides
appropriate phase-space coverage by systematically marching through action
space than by designing a grid of initial conditions $(\vx,\vv)$ for orbit
integrations.  In particular, two very different initial conditions will
specify the same orbit if an orbit passes through both phase-space points.
Consequently, when a grid of initial conditions is used to generate an orbit
library, it is not straightforward to ensure that essentially identical
orbits are not obtained at different points of the grid. When the actions are
specified, there is no danger of such double-computation. 

\item[3] When \codename\ is used, every orbit has a unique label that allows
it to be compared with orbits computed by other investigators, possibly in a
slightly different potential. 

\item[4] On account of this last point, it is trivial to
establish, indeed predetermine, the density with which the orbit library
samples phase space. This knowledge makes it possible to infer the value
taken by the \df\ on an orbit from the orbit's weight in the Schwarzschild
model. 

\item[5] When \codename\ is used, it is trivial to find with what velocities an
orbit will pass through a given point, whereas when an integrator is used,
the whole time sequence must be searched for points at which the orbit comes
near to the specified point, for in finite time it is unlikely to reach that
point exactly. This capability makes it much easier to compute velocity
distributions when \codename\ is used in place of an integrator.

\end{itemize}

The only disadvantage of using \codename\ rather than an orbit integrator is
that \codename\ cannot accurately represent resonantly trapped or chaotic
orbits. We briefly address this issue in Section~\ref{sec:resonance}, but a
full discussion of the response of \codename\ to resonant trapping lies
beyond the scope of this paper and will be the topic of a forthcoming paper.
Fortunately, in typical axisymmetric potentials, resonant trapping is of
limited significance. 

\subsection{Extensions of torus mapping}

In its simplest form torus mapping returns $(\vx,\vv)$ given $(\vtheta,\vJ)$
or $(\vv,\vtheta)$ given $(\vx,\vJ)$, but it is not well suited to computing
$(\vtheta,\vJ)$ given $(\vx,\vv)$. Consequently, much recent work has used
radically different approaches to the computation of $(\vtheta,\vJ)$ from
$(\vx,\vv)$
\citep{JJB10,JJBPJM11:dyn,JJB12:Stackel,JJB12:dfs,BoRi13,PJMJJB13,Bovy2014:streams,
SaJJB13a,Piea14_short,PifflPenoyreB,BinneyPiffl2015}.  The methods used in
these studies are inferior to torus mapping as regards precision in the sense
that they cannot be systematically refined, while by increasing the number of
coefficients $S_\vn$ used in the specification \eqref{eq:defsS} of the
generating function, torus mapping can be systematically refined. Approximate
actions often suffice for the computation of the observables of a model that
is specified by a \df\ that is an analytic function of the actions
\citep[e.g][]{JJB12:dfs,SaJJB15:Triaxial}, but lack of precision is a serious
issue when the value of the \df\ varies significantly with
changes in action comparable to the error in $\vJ(\vx,\vv)$. The classic
example of this phenomenon is the \df\ of a stellar stream
\citep{Bovy2014:streams,Sa14}.

Recently \cite{SaJJB15:Triaxial} gave an algorithm that allows one to compute
$(\vtheta,\vJ)$ from $(\vx,\vv)$ to high precision: the St\"ackel Fudge is
first used to estimate $\vJ(\vx,\vv)$. Then the torus is constructed for
$\vJ$ and the point $(\vx_1,\vv_1)$ on the torus that is nearest to
$(\vx,\vv)$ is located. Then the St\"ackel Fudge is used to estimate
$\vJ_1(\vx_1,\vv_1)$. Under the assumption that the error
$\vDelta_1=\vJ_1-\vJ$ in the actions returned by the St\"ackel Fudge is a
smooth function on phase space, an improved estimate of the actions of
$(\vx,\vv)$ is $\vJ-\vDelta_1$.  This improved estimate can be refined by
constructing the  torus $\vJ-\vDelta_1$ and finding the point $(\vx_2,\vv_2)$ on
it that is closest to $(\vx,\vv)$. Then the refined estimate of the actions
of $(\vx,\vv)$ is $\vJ-\vDelta_2=3\vJ-\vJ_1-\vJ_2$, where $\vJ_2$ are the
St\"ackel-Fudge actions of $(\vx_2,\vv_2)$ and
$\vDelta_2=\vJ_2-(\vJ-\vDelta_1)$. \cite{SaJJB15:Triaxial} showed that in
practice $\vJ-\vDelta_1$ is already an excellent estimate of the actions,
but that the algorithm can be iterated to convergence in the sense that
$(\vx,\vv)$ lies on one of the constructed tori.

In this paper we address an important topic that has been rather
neglected in recent papers on torus mapping, namely resonances. The
importance of resonances for Hamiltonian mechanics is well known
\citep{Chirikov1979,LichtenbergLieberman}. In a generic potential the
fundamental frequencies $\Omega_i$ are functions of the actions, and since
rational numbers are densely distributed on the real line, it often happens
that the frequencies are commensurable: that is, a vector $\vN$ with integer
components exists such that $\vN\cdot\vOmega=0$. If the potential has a
separable Hamilton-Jacobi equation \citep[as St\"ackel potentials do,
e.g.][]{GDII}, these resonances are accidental in the sense that they leave
no mark on the dynamics. But in general resonant orbits for which $\vN$ has
components of modest size trap neighbouring orbits, so the latter {\it
librate} around the resonant orbit, rather than circulating past it
\citep[for an account of resonant trapping, see e.g.][]{BinneyTenerife}.
Torus mapping can be used to construct an integrable Hamiltonian that is
close to the real Hamiltonian \citep{KaJJB94:MNRAS}, so trapping can be
studied with high-precision perturbation theory \citep{Ka95:closed}. In
Section~\ref{sec:resonance} we show how resonant trapping manifests itself
with torus mapping, and explain two different approaches to resonant
trapping; the approach that is appropriate depends on the astronomical
context. In Section~\ref{sec:Rzstream} we speculate that on account of
resonant trapping there may be points in the disc at which there are more
stars moving up through the plane at a given speed $v_z$ than there are
moving down at the same speed.

\subsection{Layout of the paper}

Section \ref{sec:class} is the core of the paper. It aims to demonstrate how
easily properties of individual orbits in any axisymmetric potential can be
computed with \codename, and to explain how distribution functions are used in
conjunction with tori, for example to obtain an N-body realisation of a model
galaxy. We have kept this section brief by moving as many technical details
as possible to four appendices that normal users do not need to study.
Section \ref{sec:actang} clarifies the meaning of angle variables and
presents some tests of the precision achievable with \codename.
Section~\ref{sec:interp} shows how to create tori by interpolating on a grid
of previously created tori. Section~\ref{sec:Nbody} presents a test of an
N-body realisation created with the tools included in \codename.
Section~\ref{sec:resonance} discusses the implications of resonant trapping
for torus mapping.  Section~\ref{sec:conclude} sums up and looks to the
future.

\section{Classes and namespaces}\label{sec:class}

\subsection{Units}

Internally \codename\ takes as units  $M_\odot$, kpc and Myr. In this system
Newton's constant is $G=4.99\times10^{-12}$ so for our Galaxy $GM\sim1$. The unit of
velocity, $\hbox{kpc}\Myr^{-1}=978\kms$, so the local circular speed is
$\sim0.25$ in code units. To specify a speed of $220\kms$ one writes 
{\obeylines\tt\parindent=10pt
Vc=220 * Units::kms
}
 \noindent\codename\ works with angles in radians, so to input 10 degrees one
writes
{\obeylines\tt\parindent=10pt 
theta=10 * Units::degree
}
\noindent The namespace {\tt Units} (defined in the file Units.h) contains a large range
of conversion factors between commonly used units and code units.

\begin{table}
\caption{Coordinate systems. By default the HEQ system is at epoch J2000.
$s$ denotes distance from Sun, $\mu_{\alpha*}\equiv\dot\alpha\cos\delta$ and
similarly $\mu_{l*}\equiv\dot l\cos b$. In the HCA system the $x$ axis points to the
Galactic centre and the $y$ axis points along $l=90^\circ$. In the GCA system
the $x$ axis points towards the Sun. In the GCY system the Sun is at $\phi=0$
and has $v_\phi<0$. The default local circular speed is
$\Theta_0=244.5\kms$ \citep[from the ``convenient potential'' in][]{PJM11:mass}.
The Sun's position is assumed to be  $(R,z)=(8.5,0.014)\kpc$,
and from \citet{SBD10} the Sun's
velocity in the LSR system is 
$(v_x,v_y,v_z)=(-11.1,-12.24,7.25)\kms$.}
\begin{tabular}{lcr}
Coordinate system&Point&Symbol\\
\hline
Heliocentric equatorial (J2000)&$(s,\alpha,\delta,\vlos,\mu_{\alpha*},\mu_\delta)$ &HEQ\\
Heliocentric Galactic polar&$(s,l,b,\vlos,\mu_{l*},\mu_b)$&HGP\\
Heliocentric Cartesian&$(x,y,z,v_x,v_y,v_z)$&HCA\\
Local standard of rest&$(x,y,z,v_x,v_y,v_z)$&LSR\\
Galactocentric Cartesian&$(x,y,z,v_x,v_y,v_z)$&GCA\\
Galactocentric cylindrical&$(R,z,\phi,v_R,v_z,v_\phi)$&GCY\\
\hline
\end{tabular}\label{tab:coords}
\end{table}

\subsection{Coordinate systems}

A remarkably large number of coordinate systems are useful in Galaxy
modelling, because we can consider motion to occur in the meridional $(R,z)$
plane, or in full 3d space, and for the latter we might prefer coordinates
centred on Sgr A* or on the Sun, and we might prefer to use Cartesian,
cylindrical or spherical coordinates. If we are using heliocentric polar
coordinates, we can cover the sky with either right-ascension and
declination, or Galactic longitude and latitude. Sometimes we require only
spatial coordinates, sometimes velocities, and at other times we require
phase-space coordinates. Table~\ref{tab:coords} lists the coordinate systems
available in \codename.  By defining distinct data types for each of these
numerous options, the code limits the scope for error by calling a function
with inappropriate data. Table~\ref{tab:types} lists the available data types.

\begin{table}
\caption{Data types. Types listed in the lower portion of the table are
classes that include methods for adding instances, multiplying components by
a constant, etc.}\label{tab:types}
\begin{center}
\begin{tabular}{lcr}
Type&dimension&content\\
\hline
Frequencies&3&$(\Omega_r,\Omega_z,\Omega_\phi)$\\
Actions&3&$(J_r,J_z,J_\phi)$\\
Angles&3&$(\theta_r,\theta_z,\theta_\phi)$\\
Position&3&$(R,z,\phi)$\\
Velocity&3&$(v_R,v_z,v_\phi)$\\
HEQ&6&$(s,\alpha,\delta,\vlos,\mu_{\alpha*},\mu_\delta)$\\
HGP&6&$(s,l,b,\vlos,\mu_{l*},\mu_b)$\\
HCA&6& $(x,y,z,v_x,v_y,v_z)$\\
LSR&6&$(x,y,z,v_x,v_y,v_z)$\\
GCA&6& $(x,y,z,v_x,v_y,v_z)$\\
GCY&6& $(R,z,\phi,v_R,v_z,v_\phi)$\\
\hline
PSPD&4&$(R,z,v_R,v_z)$ or\\
&& $(J_r,J_z,\theta_r,\theta_z)$\\
PSPT&6&$(R,z,\phi,v_R,v_z,v_\phi)$ or\\
&&$(J_r,J_z,J_\phi,\theta_r,\theta_z,\theta_\phi)$\\
\hline
\end{tabular}
\end{center}
\end{table}

A class {\tt OmniCoords} is defined to facilitate transformations between
coordinate systems. For example
{\obeylines\tt\parindent=10pt
OmniCoords OC; LSR w1;
w1[0]=1;w1[1]=2;w1[2]=3;
w1[3]=.01;w1[4]=.004;w1[5]=.005;
GCA w2=OC.GCAfromLSR(w1);
}
 \noindent will take the phase-space point with heliocentric location
$(x,y,z)=(1,2,3)\kpc$ and LSR velocity $(U,V,W)\simeq(10,4,5)\kms$ and return
its galactocentric position and velocity in Cartesian coordinates. The class
has methods for changing the solar motion, radius, etc., that are used in the
transformations. For example
{\obeylines\tt\parindent=10pt
OC.change\_sol\_pos(8.3,0.014); 
OC.change\_vc(220*Units::kms);
}

The default epoch is J2000, but the method {\tt change\_epoch} will change
this: for example
{\obeylines\tt\parindent10pt
OC.change\_epoch(1950)
}

\begin{table*}
\caption{Pre-defined potentials. The middle column gives the quantities fixed
by the arguments of the creator. For example, to create a log potential, write
{\tt Potential
*Phi = new LogPotential(.2,.8,.5);}}\label{tab:pot}
\begin{tabular}{lcc}
Class&arguments&formula\\
\hline
{\tt IsochronePotential}&{\tt(M,b)}&${\displaystyle {-GM\over b+\sqrt{r^2+b^2}}}$\\
 {\tt MiyamotoNagaiPotential}&{\tt(M,a,b)}&${\displaystyle \frac{-GM} {[R^2
 + (a^2 + \sqrt{b^2+z^2})^2]^{1/2}}}$\\
 {\tt LogPotential}&{\tt(V0,q,Rc)}&${\displaystyle 
\frac{V_0^2}{2}\log{\left( R^2 + \frac{z^2}{q^2} + R_c^2 \right)}
}$\\
\hline
\end{tabular}
\end{table*}

\subsection{Class Potential}

\codename\ needs to be able to evaluate a model potential and its derivatives
with respect to $R$ and $z$ -- the code is currently restricted to
axisymmetric potentials. Hence {\tt Phi(R,z)} must return the potential
value, and {\tt Phi(R,z,dPR,dPz)} must additionally leave in {\tt dPR} and
{\tt dPz} the potential's radial and vertical derivatives. Using these
derivatives, the base class provides a method $R_{\rm c}(L_z)$ that returns
the radius of the circular orbit with given angular momentum.  Since many
distribution functions involve epicycle frequencies, the class {\tt
Potential} has a method {\tt KapNuOm} that computes for a circular orbit of
given radius the radial and vertical epicycle frequencies and the azimuthal
frequency. However, tori can be constructed and manipulated without defining
this function, which requires second derivatives of the potential.

The \codename\ package includes code for the three potentials listed in
Table~\ref{tab:pot}, and for the {\sc falPot} package.  The latter computes
potentials that are generated by one or more double-exponential discs,
usually representing the thin and thick stellar discs and the gas disc, and
one or more spheroidal, double-power-law components, to represent the bulge
and dark halo. The algorithm encoded in {\sc falPot} is described by 
\cite{WDJJB98:Mass} and the code was extracted from Walter Dehnen's {\sc
falcON} package. The lines
 {\obeylines\tt\parindent=10pt
ifstream ifile; ifile.open("pot/PJM11\_best.Tpot"); 
Potential *Phi=new GalaxyPotential(ifile);
}
\noindent initialise a potential of this class. The file opened in this
example defines the ``best'' potential in \cite{PJM11:mass}, which  is
produced by thin and thick stellar discs, a flattened bulge and a
spherically symmetric dark halo. We use this ``PM11 potential'' in all our
examples. The line
{\obeylines\tt\parindent=10pt
P=(*Phi)(R,z,dPR,dPz);
}
\noindent will set {\tt P} to $\Phi(R,z)$ and leave the values of $\p\Phi/\p
R$ and $\p\Phi/\p z$ in {\tt dPR} and {\tt dPz}, respectively, and 
{\obeylines\tt\parindent=10pt
Frequencies epicycle=Phi->KapNuOm(R);
}
 \noindent will set the components of {\tt epicycle} to the frequencies
$\kappa$, $\nu$ and $\Omega$ of the circular orbit of radius $R$.

Appendix~\ref{app:pot} explains how users can define additional potentials,
and gives details of a class {\tt MultiPotential} that allows one to combine
into a single potential several previously defined potentials, for example the
potentials of a disc, a bulge and a dark halo.

\subsection{Class Torus}\label{sec:torus}

All the needs of a typical  end user of \codename\ should be covered by
the methods of the class {\tt Torus}. A torus {\tt T} with actions
$(J_r,J_z,J_\phi)=(0.1,0.2,1)\kpc^2\Myr^{-1}$ is created by the lines
{\obeylines\tt\parindent=10pt
Actions J; J[0]=.1; J[1]=.2; J[2]=1;
Torus T;
int flag = T.AutoFit(J,Phi);
}
 \noindent where {\tt Phi} is a pointer to a previously initialised model
potential.  The integer {\tt flag} returned by {\tt AutoFit} is zero if the
call was entirely successful, with values $-1$ to $-4$ signalling various
kinds of failure.  Invoked thus, {\tt AutoFit} uses default values of several
arguments that could be explicitly set. 

By far the most important of these
optional parameters is $\tolJ$, which sets the precision of the constructed
torus as follows.  \codename\ seeks to diminish the \rms\ fluctuation $\delta
H$ in $H(\vx,\vv)$ over the torus until $\delta
H<\tolJ\widetilde{\Omega}\widetilde{J}$, where
\begin{eqnarray}
\widetilde{\Omega}&\equiv&\sqrt{\Omega_r^2+\Omega_z^2},\\
\widetilde{J}&\equiv&
\begin{cases}
\sqrt{J_rJ_z}&\mbox{if }J_rJ_z\ne0\\ J_r+J_z&\mbox{otherwise.}
\end{cases}\label{eq:tildeJ}
\end{eqnarray}
One may show that when $\delta H<\tolJ\widetilde\Omega\widetilde J$, the
fluctuations in $J_r$ or $J_z$ over the constructed torus are $\la\tolJ
\widetilde{J}$.  The choice of expression (\ref{eq:tildeJ}) for the scale
action $\widetilde{J}$ is a compromise between expressions that take the
scale to be of order the larger or smaller of the two actions. This is
necessary because if, say, we have $J_r\gg J_z$ then taking
$\widetilde{J}\sim J_r$ would imply the constraint was very weak compared to
the energy that could plausibly be associated with motion in the
$z$-direction, whereas taking $\widetilde{J}\sim J_z$ would provide an
excessively stringent constraint on the accuracy required in describing the
radial motion.

Once {\tt T} has been created, it can be written to a file by

{\tt T.write\_ebf(outname,"T1,"w");}

\noindent and read back with

{\tt T.read\_ebf(outname,"T1");}

\noindent Files are written in  Sanjib Sharma's {\it Efficient Binary
Format} so \codename\ requires the  {\sc ebf} package for c++ to be
installed. It
can be found at  
sourceforge.net/projects/ebfformat/files/libebf/c-cpp/. 

\begin{figure}
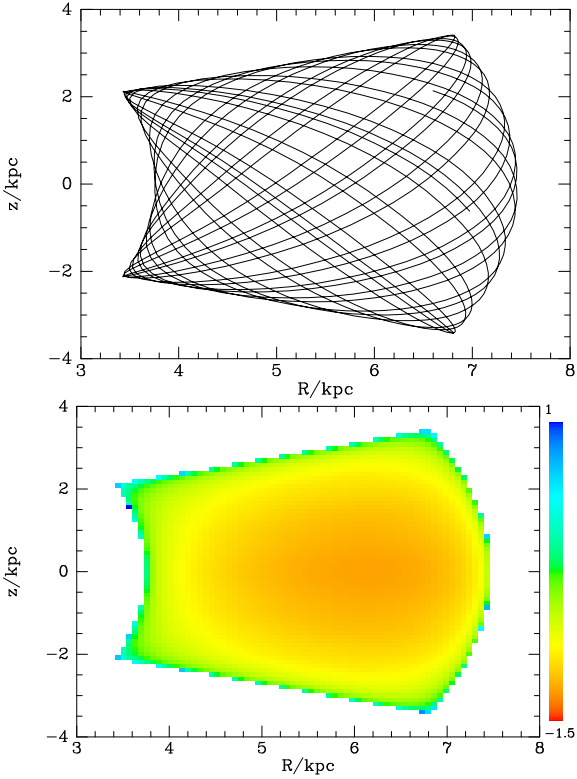

\centerline{\includegraphics[width=.9\hsize]{plots/Rz.ps}}
\centerline{\includegraphics[width=.9\hsize]{plots/rho.ps}}
\caption{Top: a time sequence along the torus
$\vJ=(0.1,0.2,1)\kpc^2\Myr^{-1}$. Bottom: logarithm of the density generated by this
torus.}
\label{fig:Rz}
\end{figure} 

The 
energy of the created torus is 
{\obeylines\tt\parindent=10pt
double E=T.energy();
}
\noindent The statements
{\obeylines\tt\parindent=10pt
Angles thetas; 
thetas[0]=1; thetas[1]=2; thetas[2]=3; 
GCY w=T.FullMap(thetas);
}
\noindent will put into the instance {\tt w} of the class {\tt GCY} of
phase-space points (Table~\ref{tab:types}) the coordinates
$(R,z,\phi,v_R,v_z,v_\phi)$ of the phase space point with the specified angle
coordinates. The lines
{\obeylines\tt\parindent=10pt
Frequencies Om; Om=T.omega();
for(int i=0; i<2048; i++)\{
\qquad thetas+=Om*2; 
\qquad w=T.FullMap(thetas);
\qquad cout << w[0] << " " << w[1] << "$\backslash$n";
\}
}
\noindent will print 2048 points $(R,z)$ along the orbit covering the elapse of
$4094\Myr$ (Top panel of Fig.~\ref{fig:Rz}).

The lines
 {\obeylines\tt\parindent=10pt
Position Rzphi; 
Rzphi[0]=3; Rzphi[1]=.4; Rzphi[2]=1; 
double s=T.DistancetoPoint(Rzphi);
}
\noindent will put {\tt s} equal to the minimum distance between the point
$(R,z,\phi)=(3,0.4,1)$ and a point on the orbit. If
the given point lies within the three-dimensional  region visited by the orbit, the returned
value of $s$ is naturally zero. Variants of
{\tt DistancetoPoint} exist that return information about the nearest
point on the torus. 

An analogous function {\tt DistancetoPSP} returns the distances in $\vx$ and
$\vv$ to the point on a torus that is closest to a given phase-space point
when the  velocity separation  is scaled to a distance using a user-specified
time-span.

 {\obeylines\tt\parindent=10pt
PSPD w4; 
w4[0]=8.5; w4[1]=0.014; w4[2]=0; w4[3]=0;
double t=4;
Vector<double,2> out=T.DistancetoPSP(w4,t,thetas);
}
\noindent will return in {\tt out[0]} and {\tt out[1]}, respectively, the real-space
and velocity-space distances of {\tt w4} from the nearest point on the
orbit, when velocities are converted to distances by multiplication by
$4\Myr$ (specified above by the value of {\tt t}). 

A method {\tt containsPoint\_Vel} is provided to discover if a given location
lies on an orbit (when 1 is returned): and, if so, find the velocities
$(v_R,v_z)$ that the star will have at that location -- two velocities
$(v_R,v_z)$ are returned, from which the four possible velocities can be
recovered by multiplying by $\pm1$:
 {\obeylines\tt\parindent=10pt
Velocity v1,v2; 
if(T.containsPoint\_Vel(Rzphi,v1,v2)==1)
printf("(\%f,\%f)(\%f,\%f)",v1[0],v1[1],v2[0],v2[1]); 
}
 \noindent An analogous method 
{\obeylines\tt\parindent=10pt
containsPoint\_Ang(Rzphi,theta1,theta2)
}
 \noindent returns the angles $(\theta_r,\theta_z)$ at which two visits occur
-- the angles of the remaining visits are $(-\theta_r,\pi-\theta_z)$.

The two methods just described call
methods that can compute more than just two velocities or two angles of a
visit. All these more sophisticated methods are called simply {\tt
containsPoint}. Some of them compute the matrix
$\p(R,z)/\p(\theta_r,\theta_z)$ and/or the determinant
$|\p(x,y,z)/\p(\theta_r,\theta_z,\theta_\phi)|$. The inverse of this
determinant, which vanishes at the edge of the orbit, is proportional to the
orbit's contribution to the density at the given point.  The bottom panel of
Fig.~\ref{fig:Rz} shows the logarithm of the density generated by the orbit
shown in the top panel. The relevant code reads
 {\obeylines\tt\parindent=10pt
	double det1,det2,**rho;
	rho=PJM::matrix<double>(NX,NX);
	for(int i=0;i<NX;i++)\{
\quad		Rzphi[0]=xmin+i*(xmax-xmin)/(float)(NX-1);
\quad		for(int j=0;j<NX;j++)\{
\quad\quad		Rzphi[1]=zmin+j*(zmax-zmin)/(float)(NX-1);
\quad\quad		int k=T.containsPoint(Rzphi,v1,det1,v2,det2);
\quad\quad			if(k==1)
\quad\quad\quad			rho[i][j]=log10(1/fabs(det1)+1/fabs(det2));
\quad\quad			else
\quad\quad\quad			rho[i][j]=-26;
\quad		\}
	\}
}

\begin{figure}
\centerline{\includegraphics[width=.8\hsize]{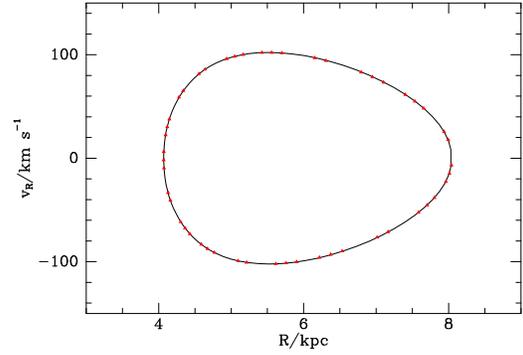}}
\caption{Full curve: surface of section of the torus
$\vJ=(0.1,0.2,1)\kpc^2\Myr^{-1}$ produced by the method {\tt SOS} of the
class {\tt Torus}. Red points: consequents along a numerically integrated
orbit.}\label{fig:SOS}
\end{figure}

Since motion in an axisymmetric potential can be reduced to motion in the
$(R,z)$ plane, surfaces of section give valuable insight into the structure
of phase space. The method {\tt SOS()} computes the $(R,v_R)$ surface of
section of a torus, so the lines
{\obeylines\tt\parindent=10pt
 	ofstream to; to.open("SOS.dat");
	T.SOS(to);
}
\noindent will write to the file {\tt SOS.dat} $(R,z,v_R,v_z,\theta_z)$ for 200 points
that have $z=0$ and $v_z>0$ (Fig.~\ref{fig:SOS}).

\subsection{Class DF}\label{sec:DF}

An abstract base class {\tt DF}  provides basic distribution functions
$f(\vJ)$. A derived class {\tt multidisk\_DF} provides \df s that are the sum
of any number of the ``quasi-isothermal'' \df s. The latter are defined by
equation (\ref{eq:totalDF}) below \citep{JJB10,JJBPJM11:dyn,PJMJJB13}. The values
taken by these \df s depend explicitly on the actions, and implicitly on the
potential through the epicycle frequencies. Instances of derived classes
store the parameters of a \df, and code that will provide the value of the
\df\ given a value of $\vJ$ and a potential.

One often needs to evaluate a \df\ at a fixed value of $\vJ$, in a fixed
potential, for many different values of the \df 's parameters. This is
required, for example, when seeking the parameter values that provide the
best fits to observational data \citep[e.g.][]{PJMJJB12,PJMJJB13}.
Significant computing time is saved by saving values, such as the epicycle
frequencies, that will be needed by all \df s. The abstract base
class {\tt quickDF} makes it possible to store the
parameters of the \df\ and any values associated with $\vJ$ and the potential
that are required to re-evaluate the \df. Instances of the derived class {\tt
multidisk\_quickDF} provide this functionality for \df s of the type
implemented by {\tt multidisk\_DF}. 

As defined by \cite{JJBPJM11:dyn}, a quasi-isothermal \df\ is
\begin{equation}\label{eq:totalDF}
  f(\vJ)\equiv{\Omega_c\nu\Sigma\over2\pi^2 M\sigma_r^2\sigma_z^2\kappa}
  \bigg|_{\Rc}
  \cut(J_\phi)\;{\rm e}^{-{\kappa J_r/\sigma_r^2}}\,{\rm e}^{-{\nu J_z/\sigma_z^2}},
\end{equation}
where $\Rc$ is the radius of a circular orbit with angular momentum
$J_\phi$, and the epicycle frequencies $\kappa$, $\nu$ and the
circular frequency $\Omega_c$ are evaluated at
$\Rc$. The choice
\[
\Sigma(J_\phi)\equiv\Sigma_{0}\e^{-\Rc/\Rd}
\]
 ensures that disc's
surface-density is an approximately exponential function of radius with
scale-length $\Rd$. $M = 2\pi\Sigma_0 \Rd^2$ is a
normalisation chosen to ensure that $(2\pi)^3\int\rd^3\vJ\,f(\vJ)=1$.
The factor $\cut(J_\phi)$ ensures we have different numbers of stars
rotating in each direction. We use the form
\[
\cut(J_\phi) = {\textstyle{1\over2}}\left[1-\tanh(J_\phi/L_0)\right],
\]
where the value of $L_0$ is small compared to the angular momentum of
the Sun.

The functions $\sigma_z(J_\phi)$ and $\sigma_r(J_\phi)$ control the
vertical and radial velocity dispersions, 
\begin{eqnarray}\label{eq:sigmas}
  \sigma_r(J_\phi)&=&\sigma_{r0}\,{\rm e}^{(R_0-\Rc)/R_\sigma}\nonumber\\
  \sigma_z(J_\phi)&=&\sigma_{z0}\,{\rm e}^{(R_0-\Rc)/R_\sigma},
\end{eqnarray}
where $R_\sigma$ is the scalelength on which the velocity dispersions
decline.

We provide a function {\tt set\_DF} that reads the parameters of a \df\ from
a file and returns a pointer to a \df. Several example input files are
included with \codename. For example 
 {\obeylines\tt\parindent=10pt
ifstream ifile; ifile.open("df/TwoDisk\_MW.df");
DF *distfunc = set\_DF(ifile);  
}
\noindent initialises a \df\ with the parameters given in the file {\tt
  TwoDisk\_MW.df}. This file reads
{\obeylines\tt\parindent=10pt
m 
2 8.5 
27 20 3.0 6.67 10 1
48 44 3.5 7.78 10 0.3
}
 \noindent where {\tt m} indicates that this is a {\tt multidisk\_DF}, {\tt 2
  8.5}
indicates that it is the sum of two quasi-isothermal \df s with
$R_0=8.5\kpc$.  The final two lines of the file give the
parameters of the two quasi-isothermal \df s 
($\sigma_{r0},\sigma_{z0},\Rd,R_\sigma,L_0,{\rm weight}$),
where the weights of the two \df s need not sum to unity, and for convenience
$\sigma_{r0}$ and $\sigma_{z0}$ are given in $\kms$, and $L_0$ in
$\kpc\kms$.

Code to find the value taken by the \df\ at a given $\vJ$ in
a given potential reads
{\obeylines\tt\parindent=10pt
  double f = distfunc->df(Phi,J);
}
 \noindent where, as ever, {\tt J} specifies the actions and {\tt Phi} points
to a {\tt Potential}.

A {\tt quickDF} cannot be set up from a file (as their raison d'etre
is that they allow for changes in the \df). They are instead set up
using a method {\tt setup}, which
takes as input (i) a pointer to a {\tt Potential}, (ii) a value $\vJ$, (iii) a
pointer to an array of parameters, (iv) a pointers to a
array of boolean values, and (v) the dimension of these arrays. For
example
{\obeylines\tt\parindent=10pt
  quickDF *qdf = new multidisk\_quickDF;
  double par[8]=\{1,8.5,0.027,0.02,3,6.67,0.01,1\}; 
%  par[0]=1; par[1]=8.5;
%  par[2]=27*Units::kms; par[3]=20*Units::kms;
%  par[4]=3.; par[5]=6.666; par[6]=10.*Units::kms;
%  par[7]=1.;
  bool changeable[8];
  for(int i=0;i<8;i++) changeable[i] = false;
  changeable[2] = true;
  qdf->setup(Phi,J,par,changeable,8);
  double fJ = qdf->df();
}
 \noindent sets up a {\tt quickDF} with essentially the same parameters as
the first of the two quasi-isothermal \df s in the {\tt DF} set up above, and
computes $f(\vJ)$. Note that for this {\tt setup} the parameters are given in
code units. The array of boolean values specifies
which parameters are liable to change: if a
parameter will not change, numbers derived from it will not need to
be recalculated. In the above example, only $\sigma_{r,0}$ is allowed to
change. To find the value of the \df\ for a different value of $\sigma_{r,0}$
one would write
 {\obeylines\tt\parindent=10pt
  par[2]=0.028; fJ = qdf->df(par);
}

\subsection{Class {\tt tunableMCMC}}\label{sec:MCMC}

N-body realisations of a particular Galaxy model can be used to generate mock
catalogues, or to initialise an N-body simulation with which to probe the
response of the model to a perturbation or its behaviour during a merger.
Instances of the class {\tt tunableMCMC} use a simple Metropolis-Hastings
Markov Chain Monte Carlo (MCMC) algorithm \citep{Metropolis} to create a
sample of $N$ actions $\vJ_i$ with associated integer weights $w_i$ such that
for any function $h(\vJ)$ of the actions
 \[
\int\rd^3\vJ\,f(\vJ)h(\vJ)\simeq\sum_iw_ih(\vJ_i)\Big/\sum_i w_i,
\]
where $f(\vJ)$ is any given \df\ normalised such that
$\int\rd^3\vJ\,f(\vJ)=1$. This MCMC algorithm avoids trying any
negative values of $J_r$ and $J_z$ by working in terms of their square
roots.

The class {\tt tunableMCMC} has methods: {\tt burn\_in}
designed to deal with burn-in of the MCMC chain; {\tt find\_sigs}
which determines the standard deviations in $\sqrt{J_r}, \sqrt{J_z}$
and $J_\phi$ to provide a sensible proposal density for the MCMC chain
(a Gaussian with these standard deviations in each direction); {\tt
tune} to tune the step size to a more appropriate value in each
direction; and {\tt output} which gives the output from the chain,
either to a file or to vectors.

Included with \codename\ are three main programs that perform the steps
required to create an N-body model from a \df. {\tt Choose\_any\_df}
outputs a list of actions and their associated weights sampled from a {\tt
DF} using {\tt tunableMCMC}. {\tt Create\_df\_tori} takes this list
and creates a torus for each one, storing it and its
weight in a file. {\tt Sample\_list\_limits} takes this file and
outputs points in Galactocentric Cartesian coordinates. These points are
uniformly distributed in $\vtheta$ over tori that are
randomly selected from the previously written list using the weights $w_i$. 

Code that does a similar job, although omitting output at intermediate
steps and sampling the tori in a random order, is as follows

{\obeylines\tt\parindent=10pt
  time\_t cpu=time(NULL);
  Random3   R3(7*int(cpu)),R3b(123*int(cpu));
  Gaussian  Gau(\&R3,\&R3b);
  J[0] = J[1] = 0.001; J[2] = -1.8; 
  tunableMCMC tMC(distfunc,Phi,J,\&Gau,\&R3);
  int nTor=3000, nEach=10;
  tMC.burn\_in(nTor/10);  
  tMC.find\_sigs(nTor/10);
  tMC.tune(nTor/10,2);
  vector<Actions> Jtab; vector<int> wtab;
  tMC.output(nTor,\&Jtab,\&wtab);
  for(int i=0;i<nTor;i++)\{
\quad    T.AutoFit(Jtab[i],Phi);
\quad    for(int j=0;j<nEach*wtab[i];j++)\{
%\qquad	    Angles A;
\qquad      for(int k=0;k<3;k++) 
\qquad\qquad thetas[k] = 2*Pi*R3.RandomDouble(); 
\qquad      cout << T.FullMap(thetas) << '$\backslash$n';
\quad\}
\}
}

%%%%%%%%%%%%%%%%%%%%%%%%%%%%%%
\section{Angle coordinates} \label{sec:actang}

In principle any point on the torus may be assigned $\vtheta=0$, but there
are natural choices for the zero point. \codename\ takes the zero point to be
where the star is at pericentre with $z=0$ and at azimuth $\phi=0$.
Given this zero point,
there are some useful rules of thumb for interpreting angle
coordinates.
\begin{itemize}
\item The value $\theta_\phi$ is  the $\phi$
  coordinate of the guiding centre of the star's epicycles.
  
\item $\theta_r\sim0$ corresponds to pericentre, and $\theta_r\sim\pi$
  to apocentre. We therefore typically have $v_R>0$ for 
  $0\lesssim\theta_r\lesssim\pi$, and $v_R<0$ for 
  $\pi\la\theta_r\lesssim2\pi$.
  
\item The orbit passes upwards through $z=0$ at $\theta_z\sim0$,
  reaches maximum distance above the plane ($z>0$) at
  $\theta_z\sim\pi/2$, passes through the plane again (now with
  $v_z<0$) at $\theta_z\sim\pi$, then reaches maximum distance below
  the plane at $\theta_z\sim3\pi/2$.
\end{itemize}

\begin{figure}
  \centerline{\includegraphics[width=.85\hsize]{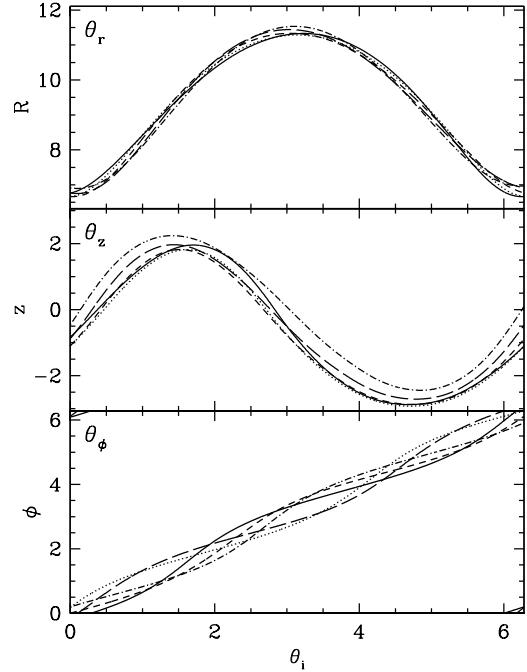}}
  \caption{Typical relations between $\vtheta$
    and the real-space position of a thick-disc star. Each line shows how a
    real-space variable changes with its associated  angle variable with the
    other two angle variables held constant. Different lines arise from
    different values of the constant angles. We see that $\phi\approx\theta_\phi$,
    $z\sim\sin\theta_z$ and $R-R_{\rm g}\sim -\cos\theta_r$, where
    $R_{\rm g}$ is the guiding centre radius. In the limit $J_r, J_z
    \rightarrow 0$ these approximations become exact.
\label{fig:th_coords}
}
\end{figure}

\noindent These relations
would be exact if the motion were separable in $R$ and
$z$. Fig.~\ref{fig:th_coords} shows a typical case.

\begin{figure*}
  \centerline{\resizebox{0.5\hsize}{!}{\includegraphics[angle=270]{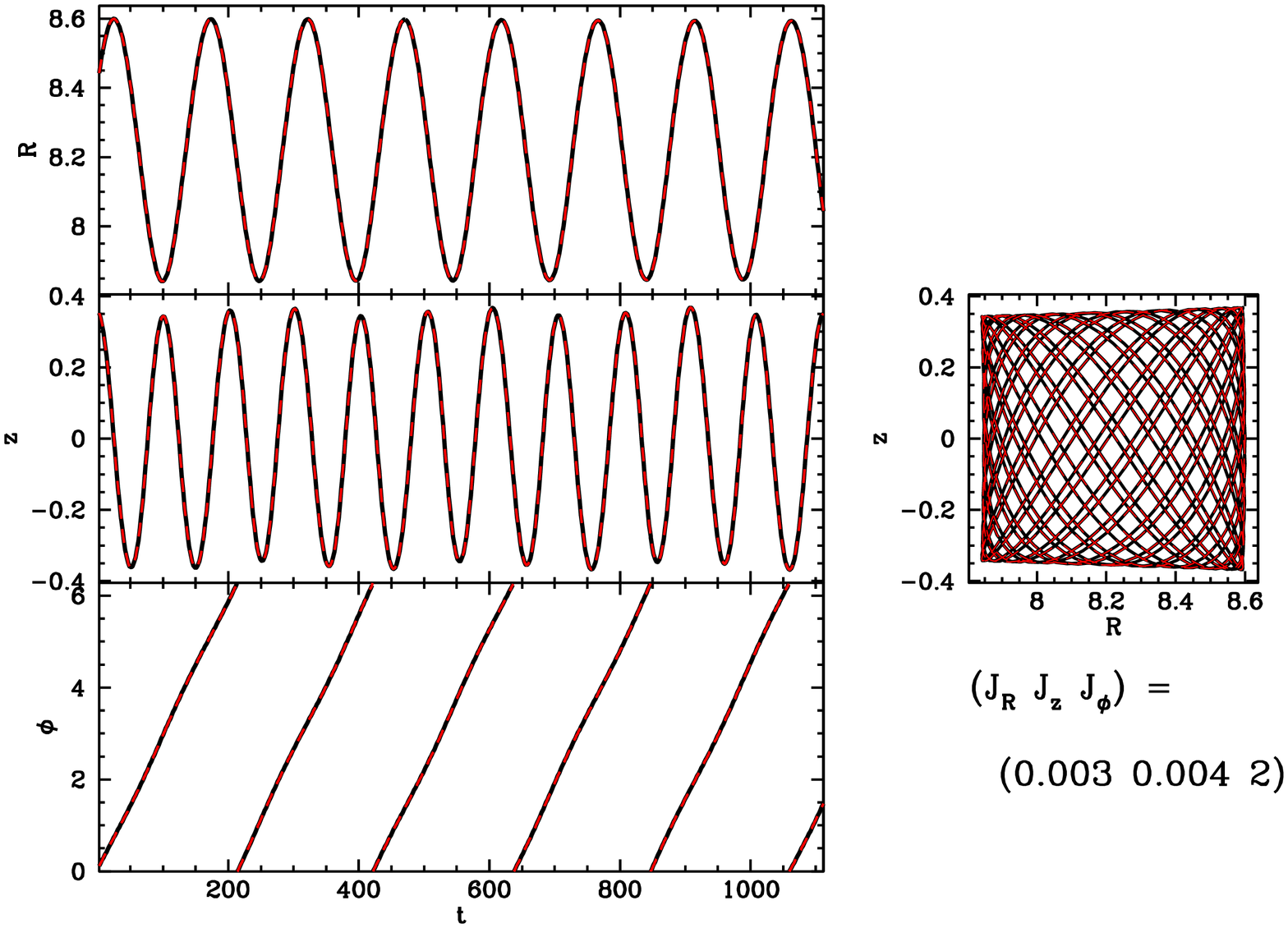}}
    \resizebox{0.5\hsize}{!}{\includegraphics[angle=270]{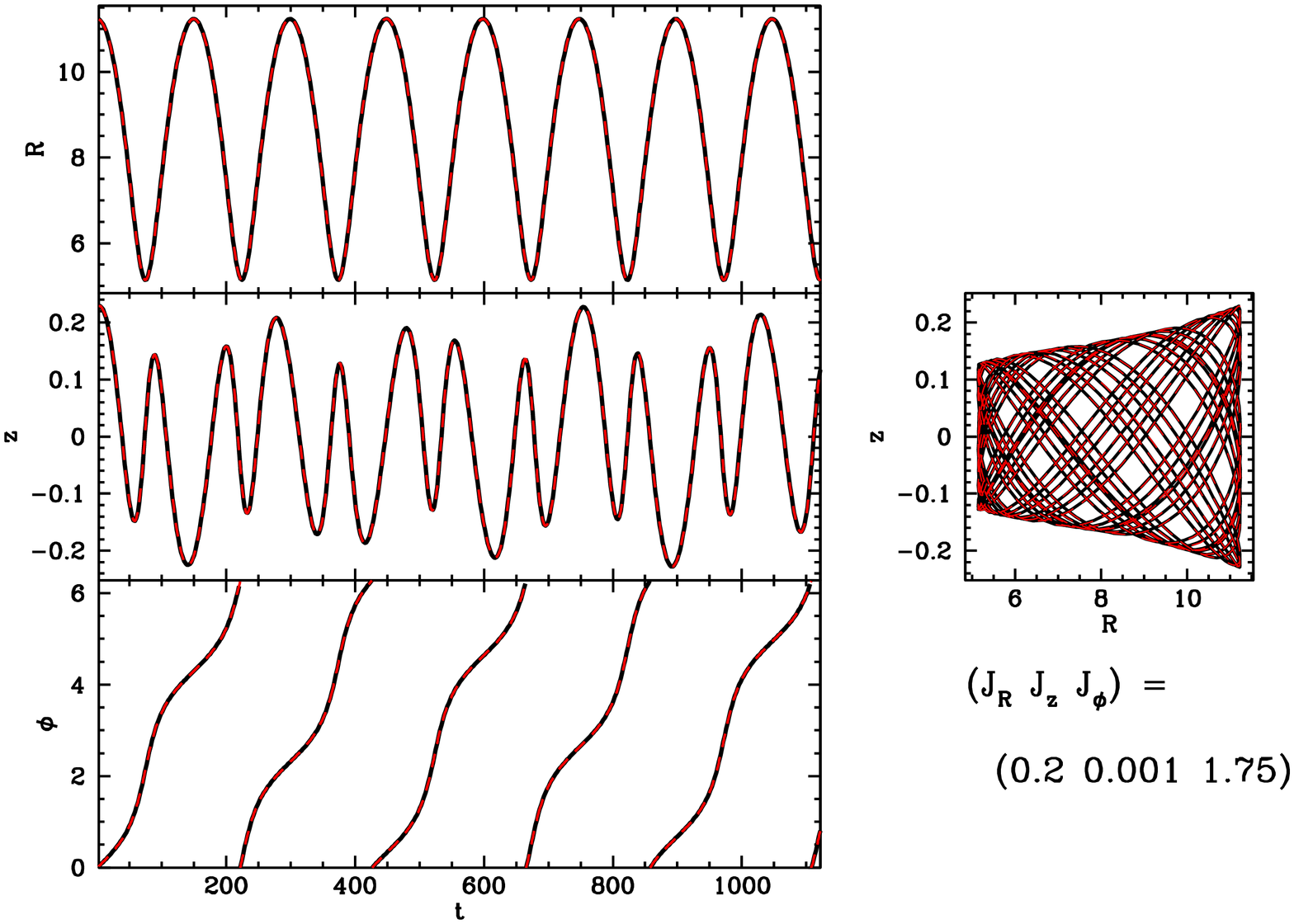}}}
  \vspace{-1mm}
  \centerline{\resizebox{0.5\hsize}{!}{\includegraphics[angle=270]{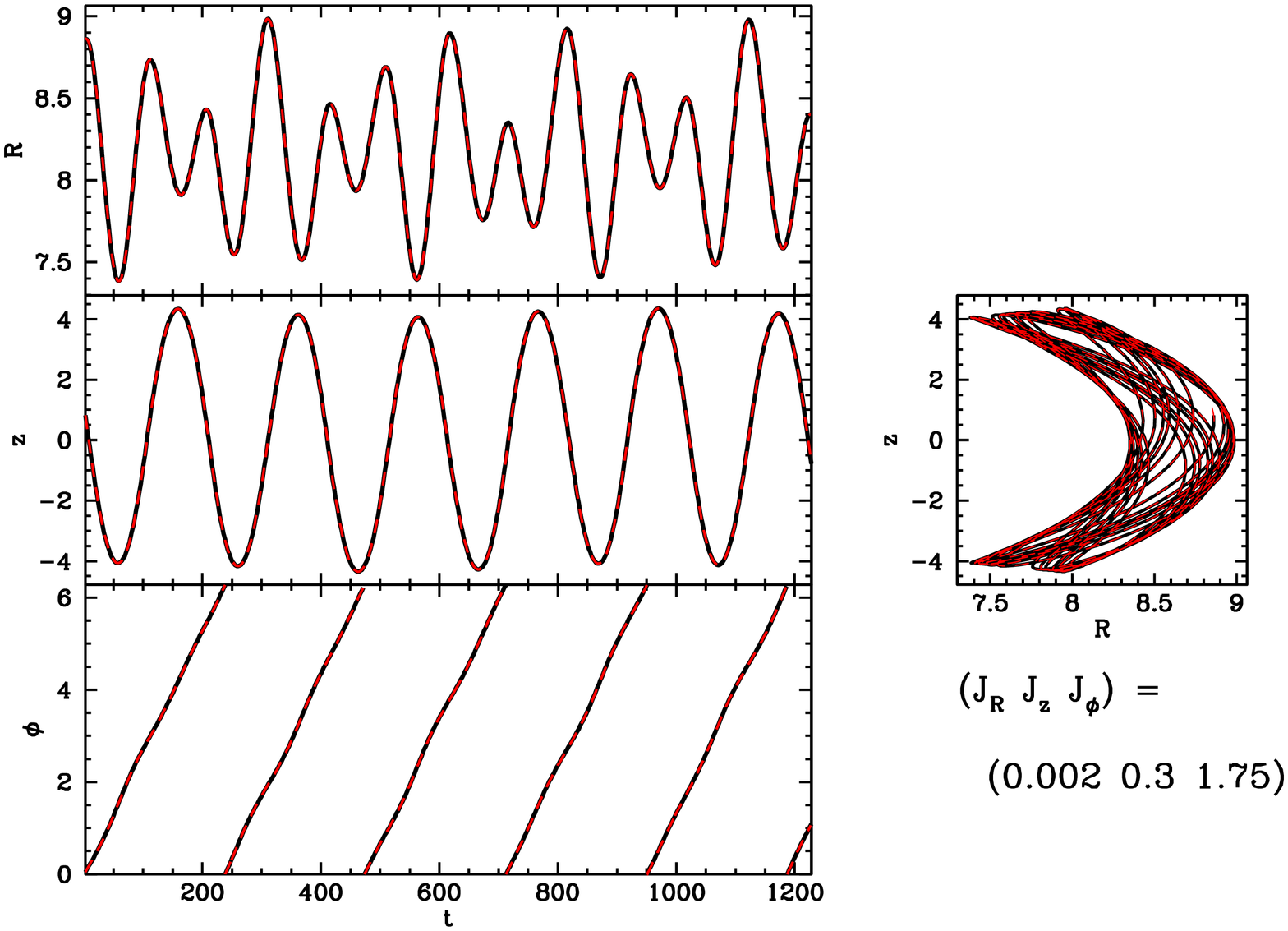}}
    \resizebox{0.5\hsize}{!}{\includegraphics[angle=270]{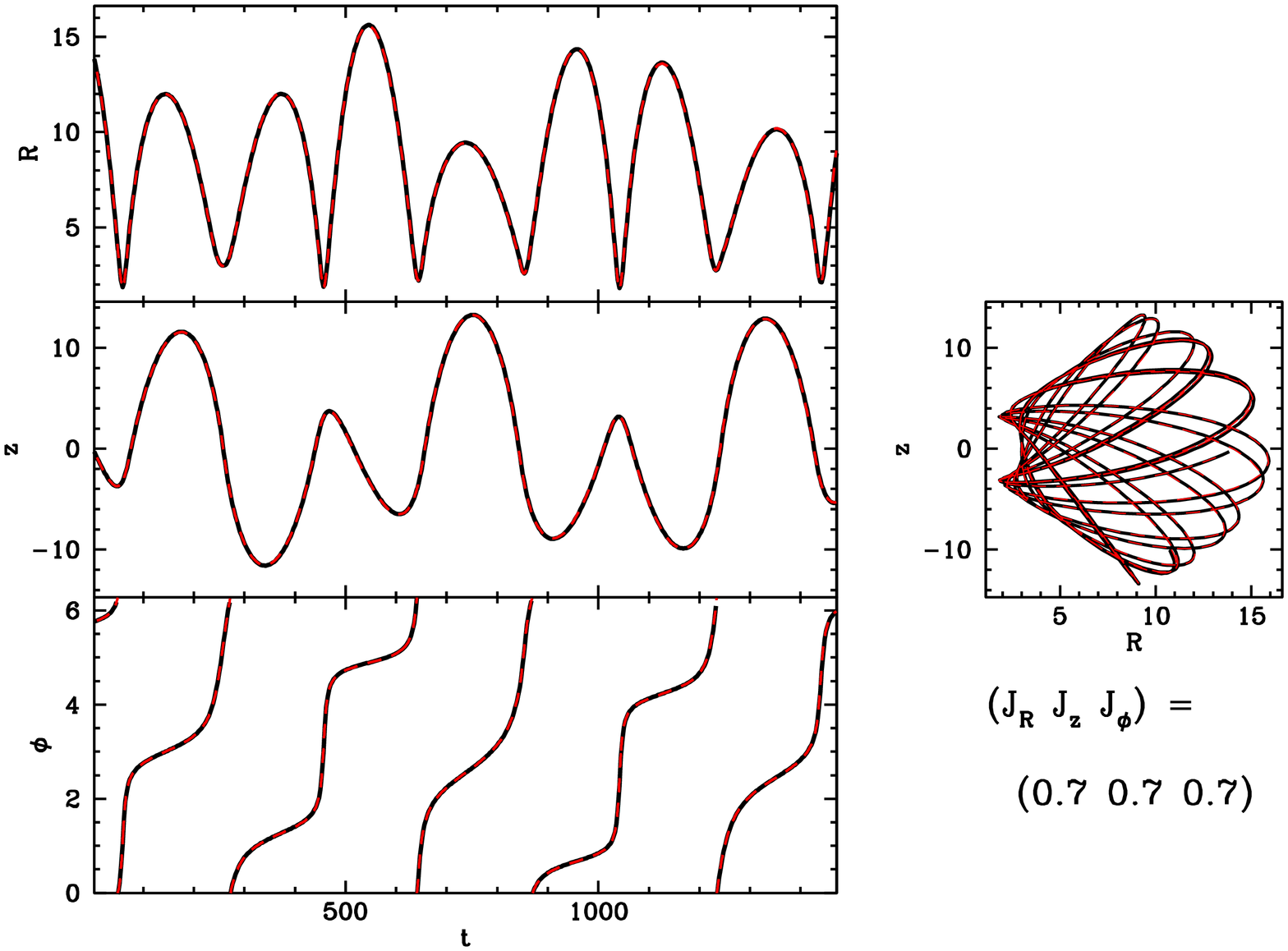}}}
  \caption{
    Orbits found by tracing the path associated with the expected
    linear increase in $\vtheta$ using the torus machinery (solid
    black lines) or by Runge-Kutta integration in the gravitational
    potential (red dashed lines). The two agree to high precision.
\label{fig:orbit}
}
\end{figure*}

The requirement that $\vtheta$ increase linearly in time along an orbit
computed using a Runge-Kutta or similar integrator provides a rigorous
test of the accuracy of fitted tori.  Fig.~\ref{fig:orbit} shows this test
for four tori. At top left we have a
nearly circular orbit, at top right an  orbit that is quite eccentric but  lies
nearly in the equatorial plane, at bottom left a less eccentric orbit that
moves far from the plane, and at bottom right an eccentric and highly
inclined orbit such as might be occupied by a halo globular cluster.
Paths computed using the tori are shown by solid black lines and those
computed by Runge-Kutta are shown with dashed red lines. The near coincidence
of these lines demonstrates the precision that \codename\ achieves.

All four tori were fitted with tolerance threshold $\tolJ=0.0002$, which is
significantly smaller than we typically use, in order to demonstrates the
accuracy with which torus mapping can reproduce orbits that are far from
resonance. As we explain below, the phenomenon of resonant trapping makes it
expedient to employ the larger value $\tolJ=0.003$ in general.

\section{Interpolating between tori}\label{sec:interp}

In some applications one needs to sample densely a small region of action
space -- a prime example is when modelling stellar streams
\citep{Sa14,Bovy2014:streams}.
Dense sampling of action space is also required when implementing Hamiltonian
perturbation theory using an integrable Hamiltonian defined by tori
\citep{Ka95:closed}.  When
dense sampling is required, tori are best obtained by interpolation on a grid
of fitted tori rather than by fitting all tori directly to the Hamiltonian.
Moreover, we will see in Section~\ref{sec:resonance} that when resonant
trapping is important, it is absolutely essential to obtain tori for certain
``missing actions'' by interpolating between tori obtained for actions that
are not subject to trapping.

Interpolation between tori was introduced by \cite{KaJJB94:PhysRev} to
implement Hamiltonian perturbation theory. They interpolated the Fourier
coefficients $S_\vn$ of the generating function (eqn.~\ref{eq:defsS}), their
derivatives with respect to $\vJ$, and the parameters of the toy potential
$\PhiT$.
\codename\ implements this scheme by making it possible to multiply tori by
any real number and add them: these operations are interpreted as the
corresponding operations on each $S_\vn$ and each parameter of $\PhiT$.
\cite{KaJJB94:PhysRev}
showed that linear interpolation in $\vJ$ provides an acceptable
approximation, and we employ this method here, although in principle a
higher-order interpolation scheme could be developed that takes advantage of
the values of the derivatives $\p S_\vn/\p\vJ$ for each fitted torus.

\begin{figure}
  \centerline{{\includegraphics[width=.83\hsize,angle=270]{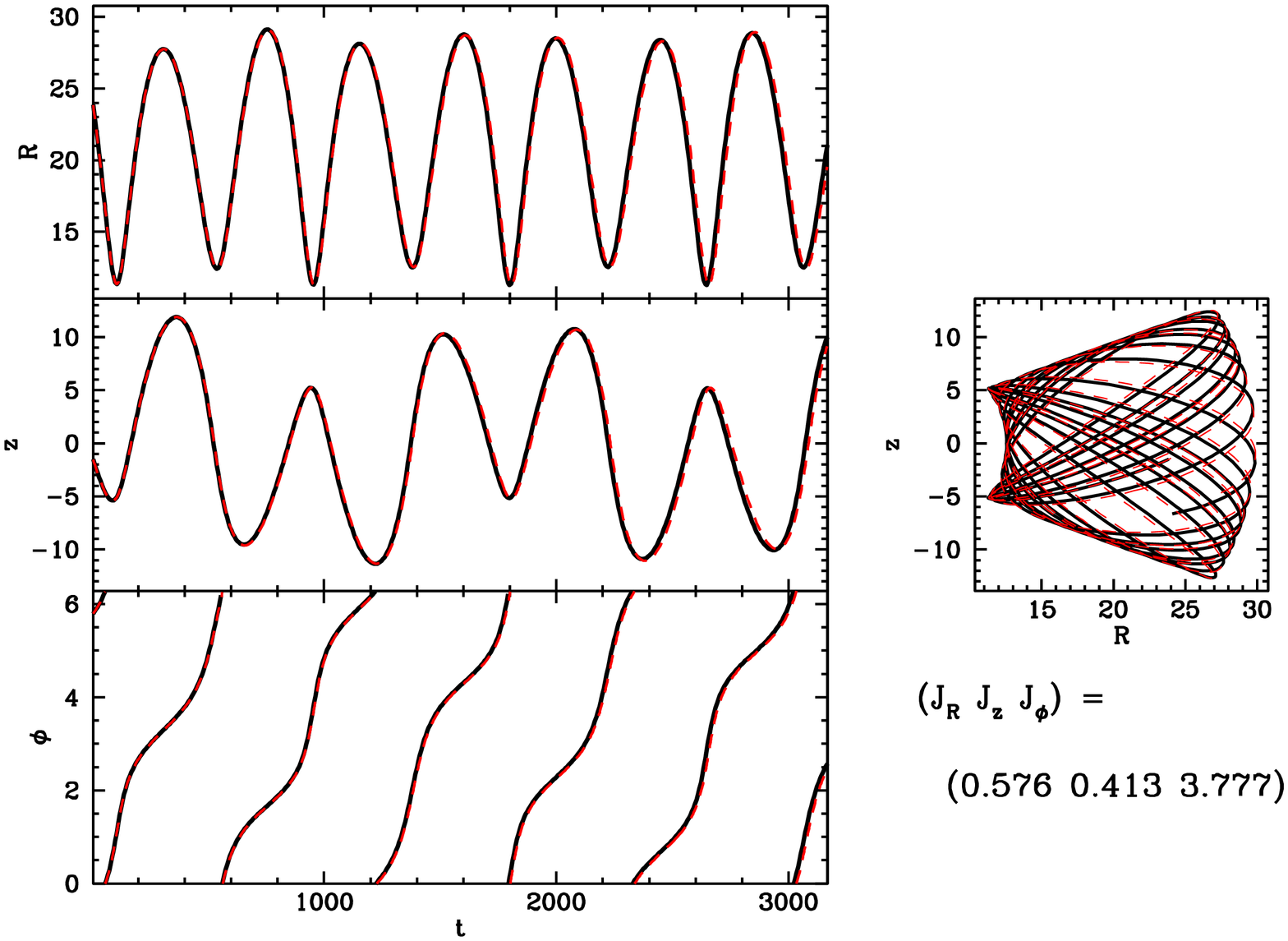}}}
  \vspace{-1mm}
  \centerline{{\includegraphics[width=.83\hsize,angle=270]{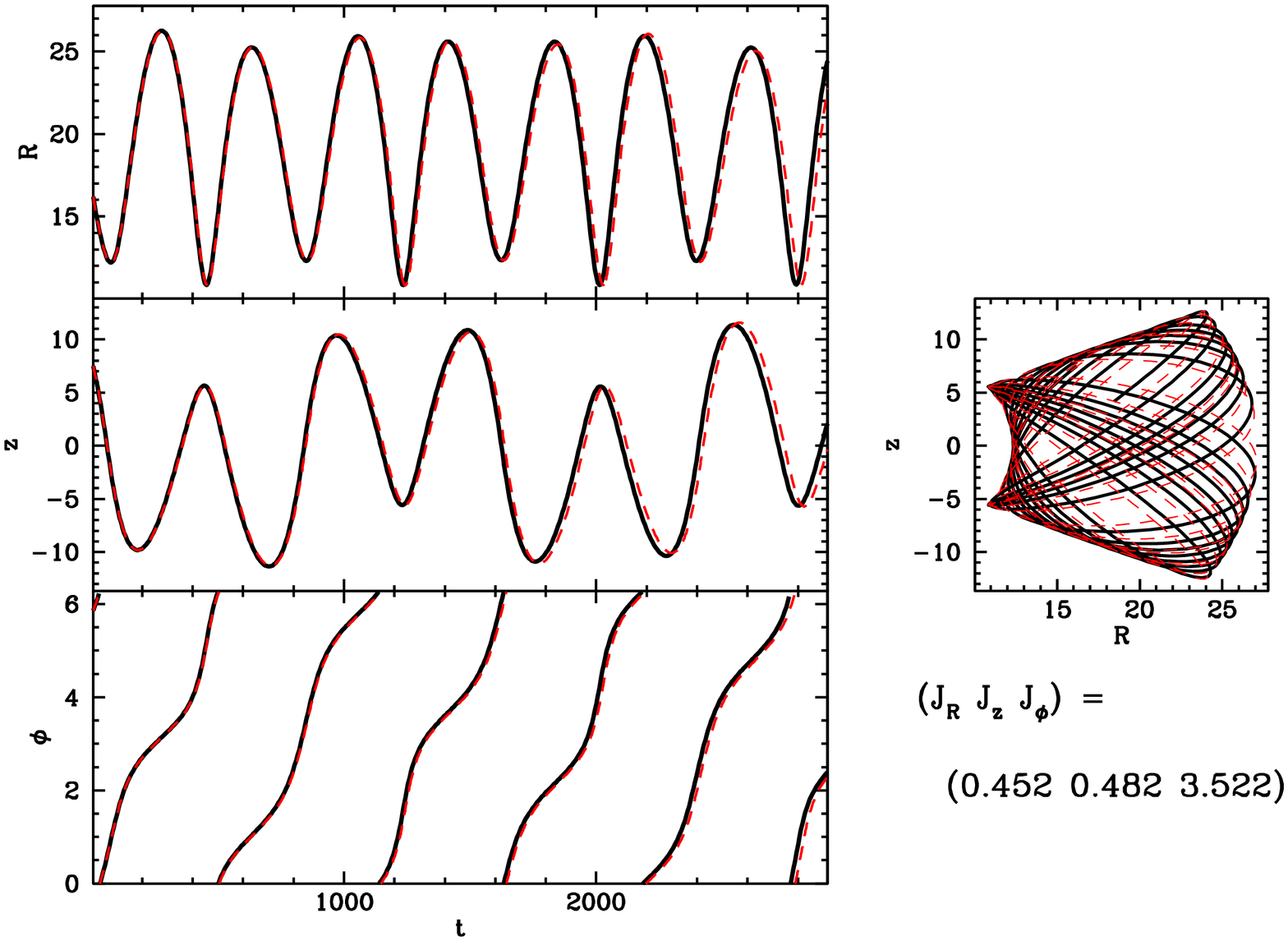}}}
  \caption{
    Orbits found by tracing the path associated with the expected
    linear increase in $\vtheta$ using interpolated tori (solid
    black lines) or by Runge-Kutta integration in the gravitational
    potential (red dashed lines). 
\label{fig:checkinterpolate}
}
\end{figure}

Code to enable the construction of  tori near the torus {\tt T} reads
{\obeylines\tt\parindent=10pt
	Actions Jbar=T.actions(),dJ;
	for(int j=0;j<3;j++) dJ[j]=.03;
	Torus Tg; Actions Jg;
	Torus ***Tgrid = PJM::matrix<Torus>(2,2,2);
	for(int i=0;i<2;i++)\{
\qquad		Jg[0]=Jbar[0]+(i-.5)*dJ[0];
\qquad		for(int j=0;j<2;j++)\{
\qquad\qquad		Jg[1]=Jbar[1]+(j-.5)*dJ[1];
\qquad\qquad		for(int k=0;k<2;k++)\{
\qquad\qquad\qquad		Jg[2]=Jbar[2]+(k-.5)*dJ[2];
\qquad\qquad\qquad		Tg.SetToyPot(Phi,Jg);
\qquad\qquad\qquad		Tgrid[i][j][k].FitWithFixToyPot
\qquad\qquad\qquad\quad (Jg,Tg.TP(),Phi,.001);
\qquad\qquad		\}
\qquad		\}
	\}
	Torus T2=InterpTorus(Tgrid,Jbar,dJ,Jbar);
}
 \noindent Here we first declare an array of eight tori arranged at the
corners of a cube in action space surrounding {\tt T} and
$0.1\kpc^2\Myr^{-1}$ on a side.  Then we create a torus at each grid point by
a two-step process: {\tt SetToyPot} finds the parameters of a suitable toy
potential, and then {\tt FitWithFixToyPot} optimises the $S_\vn$ for this
$\PhiT$. Finally by interpolation on the corners of the cube we create a
torus {\tt T2} that should be very close to the original torus {\tt
T}.\footnote{Normally tori are fitted by {\tt AutoFit}, which adjusts the toy
potential in parallel with the $S_\vn$. Here we use {\tt FitWithFixToyPot},
which fits only the $S_\vn$ because when $\PhiT$ is fitted simultaneously
with the $S_\vn$, trade-offs between them
tend to impair smooth variation of the parameters with $\vJ$.}
Analogously to Fig.~\ref{fig:orbit}, Fig.~\ref{fig:checkinterpolate} shows
comparisons of integrated orbits and time sequences over tori constructed by
this interpolation procedure. The grid was formed by the corners of a cuboid
in action space centred on $\overline{\vJ}=(0.5,0.5,3.55)\kpc^2\Myr^{-1}$.
The grid's diagonal is $\Delta\vJ=(0.2,0.2,0.7)\kpc^2\Myr^{-1}$. The actions at
which two tori were obtained were randomly chosen from within the cuboid and
are given below each right-hand panel of Fig.~\ref{fig:checkinterpolate}. The
dashed red lines  show the result of
Runge-Kutta integration, while the black lines show time sequences along two
tori obtained by interpolation. The curves
overlie each other to a satisfactory extent.

An independent check on the accuracy of an interpolated torus is provided by the
\rms\  variation $\delta H$ of the Hamiltonian (\ref{eq:defsH}) over
the torus. The grid tori were fitted with tolerance parameter $\tolJ=0.001$,
which gives us an average $\delta H=2\times10^{-3}\kpc^2\Myr^{-2}$ for these
tori. The average interpolated torus has $\delta
H=4\times10^{-3}\kpc^2\Myr^{-2}$, which indicates perfectly acceptable
accuracy.

\begin{figure}
  \centerline{\resizebox{\hsize}{!}{\includegraphics{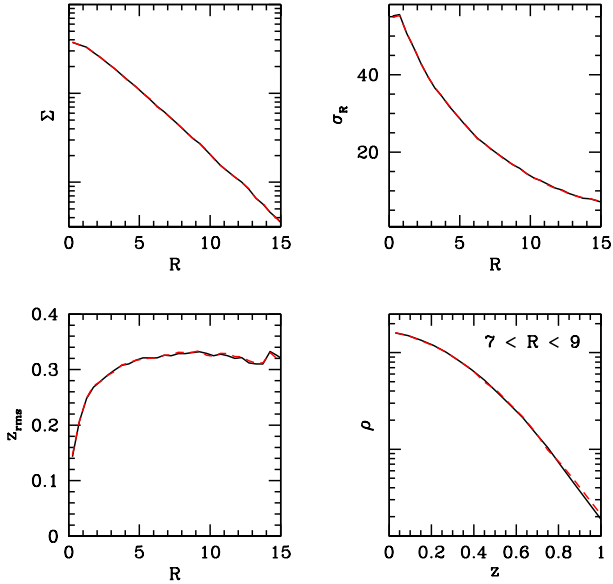}}}
  \caption{Properties of an $f(\vJ)$ model sampled by \codename\
    (solid black line) and the same model evolved by Runge-Kutta
    integration of the orbits in the Galactic potential for $5\Gyr$
    (red dashed line). The figures show the surface density $\Sigma$ (top
    left), radial velocity dispersion $\sigma_R$ (top right) and 
    root-mean-squared distance from the plane $z_{\rm rms}$ (bottom left)
    as a function of Galactocentric radius, and the density profile
    $\rho$ as a function of $z$ for particles with Galactocentric
    radius $7\kpc < R < 9\kpc$. The lines are almost identical because
    the model starts from statistical equilibrium, as it should.
    \label{fig:fJ}
  }
\end{figure}

In a
future paper we will use torus interpolation to model  stellar
streams.

 \section{N-body models} \label{sec:Nbody}

As described in Section~\ref{sec:MCMC}, \codename\ provides a means of
choosing initial conditions for N-body simulations of realistic disc galaxies
that start in perfect equilibrium.  Here we illustrate this technique in the
case of a tracer population that moves in a fixed potential. However, now
that \cite{BinneyPiffl2015} have shown how to determine the potential that
is jointly created by populations of stars and dark-matter particles that
each have prescribed \df s $f(\vJ)$, one could in principle use tori to set
up the initial conditions of a self-consistent galaxy by first relaxing the
joint potential as \cite{BinneyPiffl2015} describe and then using a pointer
{\tt Phi} to that potential. Here {\tt Phi} points to the PM11 potential.

Fig.~\ref{fig:fJ} shows plots for a population sampled from a \df\ of the
form (\ref{eq:totalDF}). The parameters of the \df\ are $\Rd=3\kpc$,
$R_\sigma=6.67\kpc$, $\sigma_{r0}=\sigma_{z0}=20\kms$ and $L_0=10\kpc\kms$.
$50\,000$ values of $\vJ$ were sampled from this \df\ using an instance of
{\tt tunableMCMC}.\footnote{As explained in Section~\ref{sec:MCMC}, some
values of $\vJ$ occur $w>1$ times in the MC chain, and in this case $\vJ$ is
specified only once but given weight $w>1$. We sample  $50\,000$ distinct
tori, some with weights $w_i>1$. Then the probability that the $i$th torus will be
drawn  from this library is $w_i/\sum_jw_j$, and on the selected torus 20
points are chosen.} We find the corresponding tori, and from each torus sample 20 values of
$\vtheta$, so we have $10^6$ particles in all. On a single core of a typical
laptop it takes $90$ minutes to fit the $50\,000$ tori (a rate of
$\sim10\;{\rm s}^{-1}$) and $2$ minutes to sample the $10^6$ particles (a
rate of $\sim10\,000\;{\rm s}^{-1}$). Both tasks are trivially
parallelisable.

The full black lines in Fig.~\ref{fig:fJ} show the surface density, radial
and vertical velocity dispersions, disk thickness and vertical profile in the
range $7\kpc<R<9\kpc$ that result from the sampling. The dashed red lines
show the same profiles after the orbit of every particle has been evolved
with a Runge-Kutta integrator for $5\Gyr$. The differences between the
initial and evolved distributions are tiny and consistent with Poisson noise.
This demonstrates that the \df\ is has been properly sampled, so the initial
distribution is consistent with the Jeans theorem.

\section{Resonances} \label{sec:resonance}

Real galaxy potentials are almost certainly not integrable. That is, these
potentials manifest the phenomenon of resonant trapping, in which orbits
librate around an orbit with commensurable fundamental frequencies, so
$\vN\cdot\vOmega=0$ for a vector $\vN$ with integer components. The extent of
this phenomenon varies widely from potential to potential, for reasons that
are not fully understood \citep[see][\S3.7.3 for an interesting
example]{GDII}. 

When orbits are effectively two-dimensional (as in an axisymmetric
potential), the extent of resonant trapping can be determined from surfaces
of section. Resonant trapping of fully three-dimensional orbits is best
probed with a frequency map \citep[\S3.7.3(b)]{Laskar1993,GDII}.  In general
the fraction of phase space occupied by trapped orbits is small in typical
axisymmetric potentials, and moderate in triaxial potentials that have large
core radii and low pattern speeds. Shrinking the core radius of a triaxial
potential increases the fraction of trapped orbits
\citep{MerrittValluri1999}, and even with a large core radius, orbits
comparable in size to the potential's corotation radius have a significant
probability of being resonantly trapped.

Resonantly trapped orbits are quasiperiodic like their untrapped cousins. But
when the regions of influence of individual resonances overlap, orbits cease
to be quasiperiodic \citep{Chirikov1979} and one says that they are
chaotic. Numerical experiments suggest that many chaotic orbits are
successively trapped by different resonances, with the consequence that their
frequencies are stable for only brief periods of time.

When we use a torus-mapping code, we pre-determine the gross structure of the
image torus by specifying the toy potential and any point transformation, and
the code merely tweaks the image torus to make $H$ as constant as possible
over its surface. 

In the case of \codename, the tori have the gross structure of non-resonant
tori. When such a non-resonant orbital torus exists with the specified
actions, \codename\ should in principle be able to drive $\delta H$ to
arbitrarily small values. When there is no non-resonant torus with the
specified actions, it is impossible for \codename\ to drive $\delta H$ to
zero. Hence when \codename\ is employed, resonant trapping can be signalled
by failure of {\tt AutoFit} to achieve a small specified value of $\tolJ$ and
consequently return a non-zero value.

\begin{figure*}
 \centerline{\includegraphics[width=3.75cm,height=3.75cm]{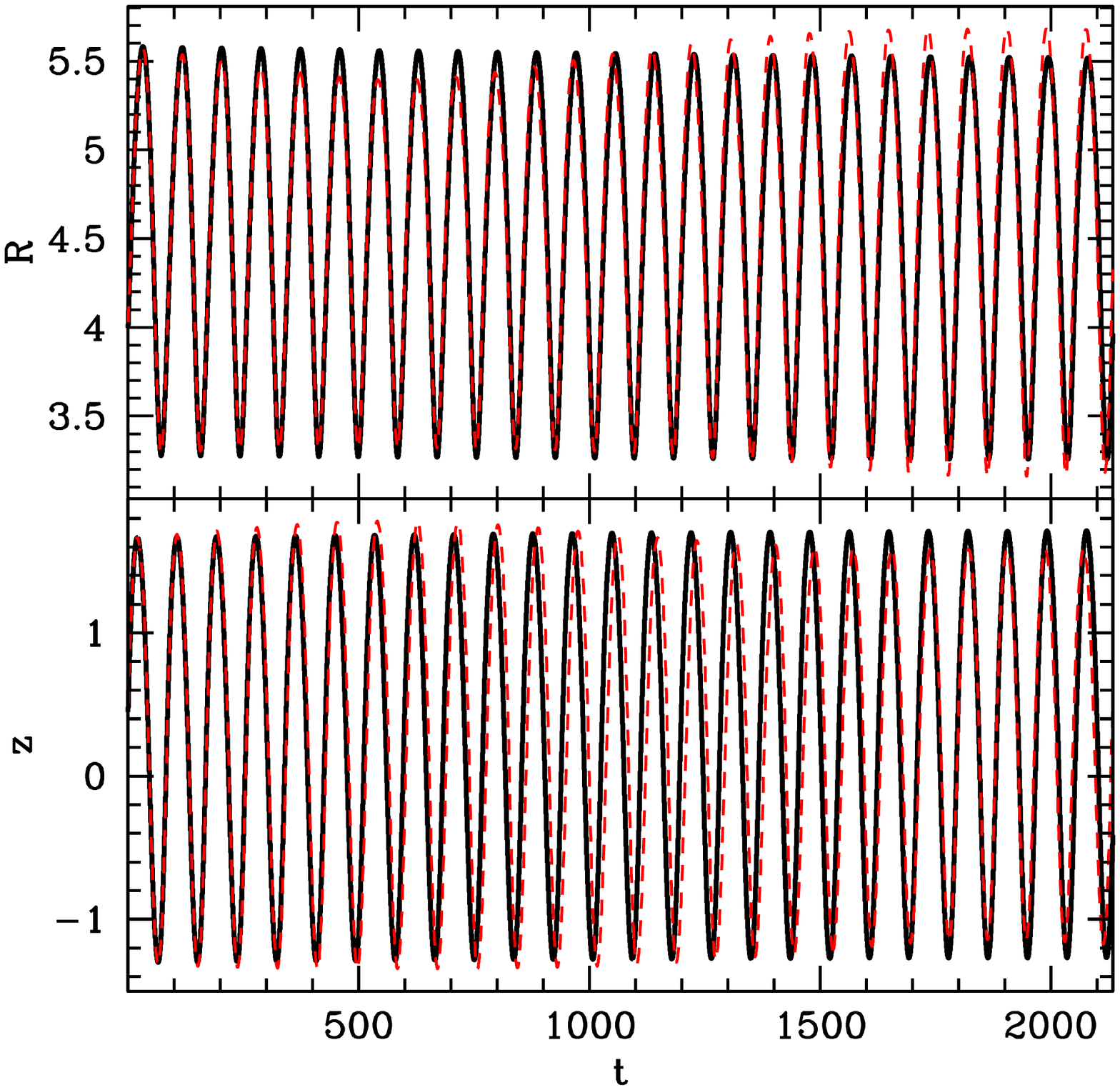}
    \includegraphics[width=3.75cm,height=3.65cm]{plots/res_top.sos2.ps}
    \includegraphics[width=3.75cm,height=3.65cm]{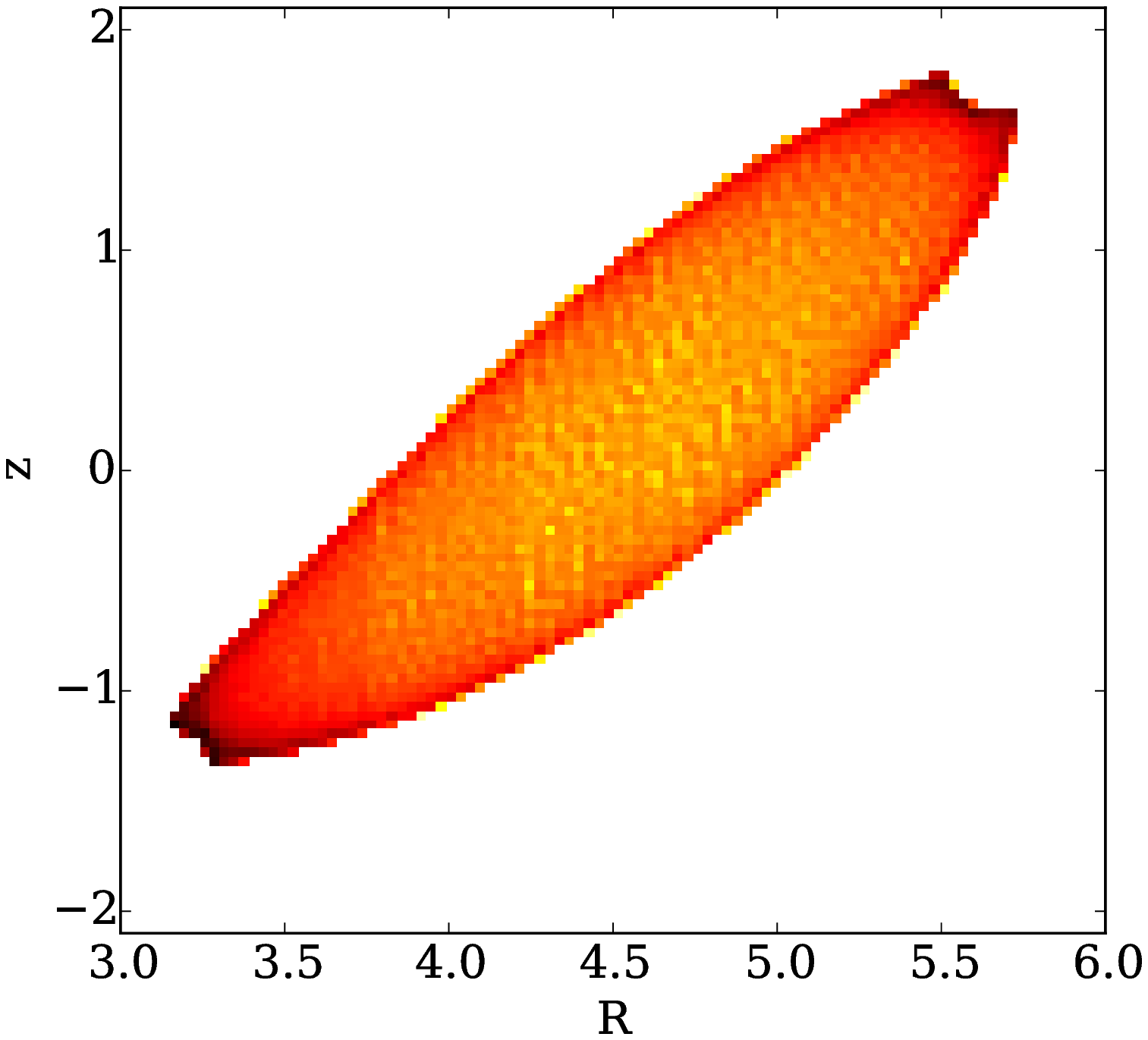}
    \includegraphics[width=3.75cm,height=3.65cm]{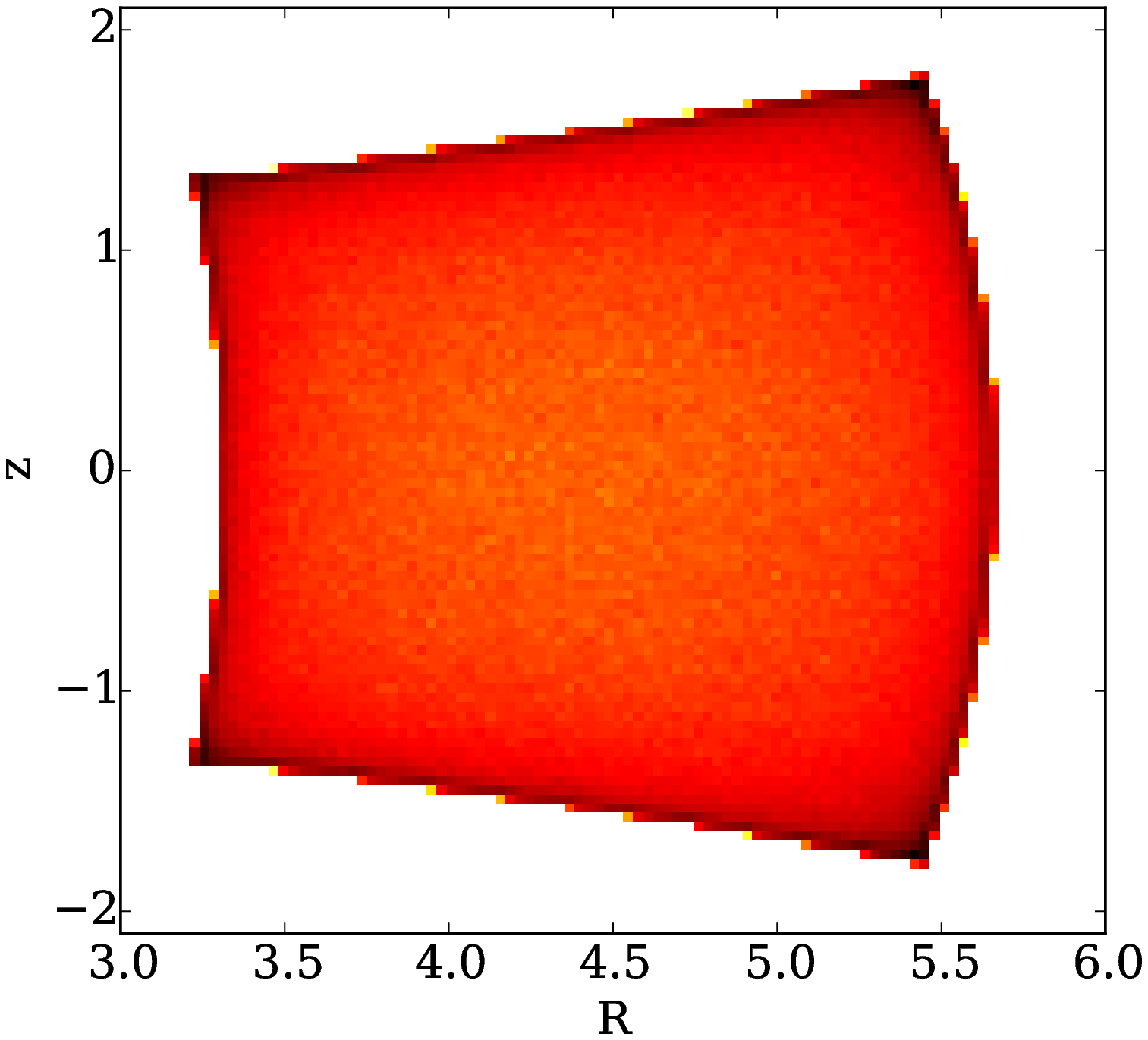}}
 \centerline{\resizebox{3.75cm}{!}{\includegraphics{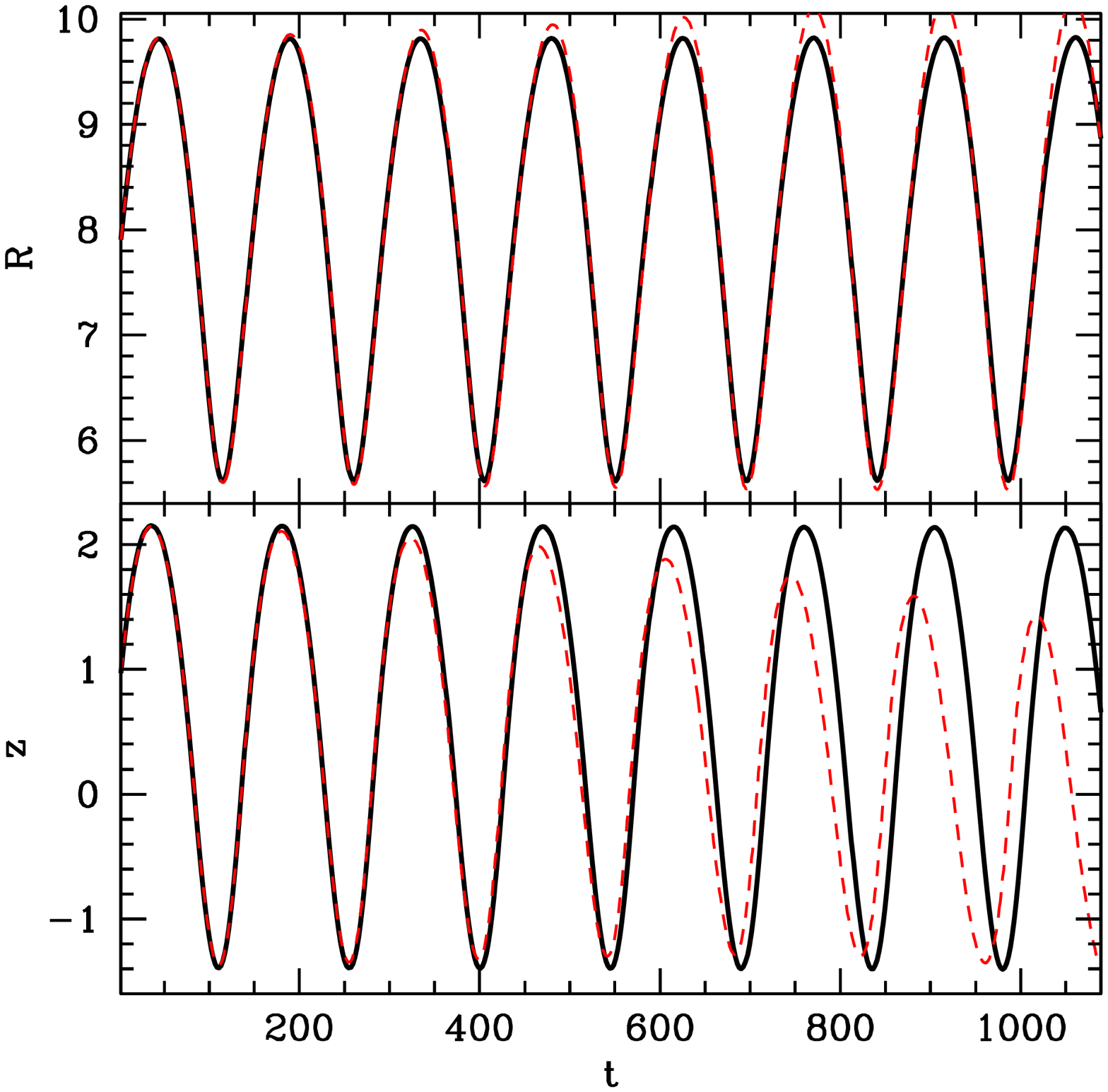}}
    \resizebox{3.75cm}{!}{\includegraphics{plots/res_front.sos2.ps}}
    \resizebox{3.75cm}{!}{\includegraphics{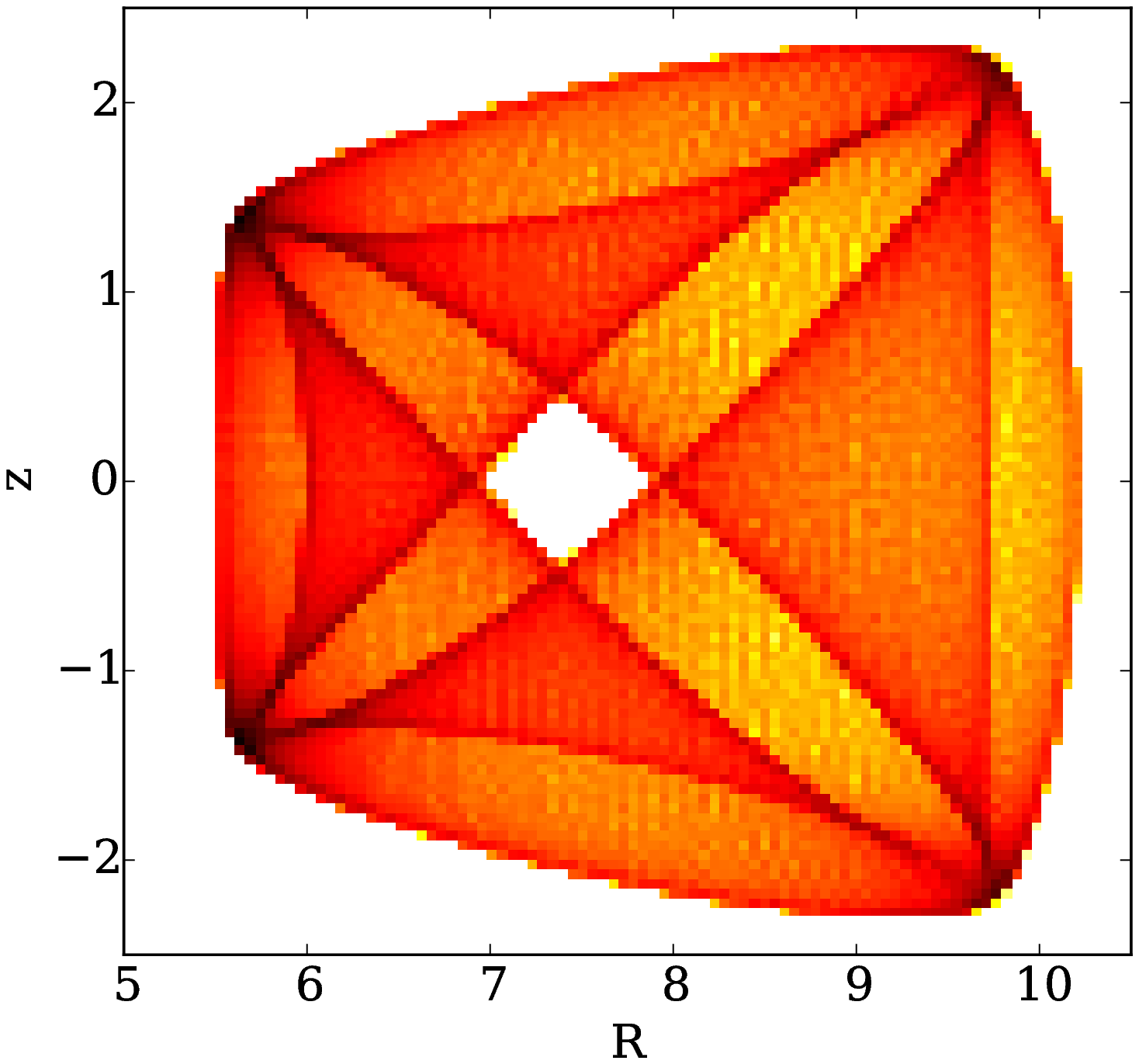}}
    \resizebox{3.75cm}{!}{\includegraphics{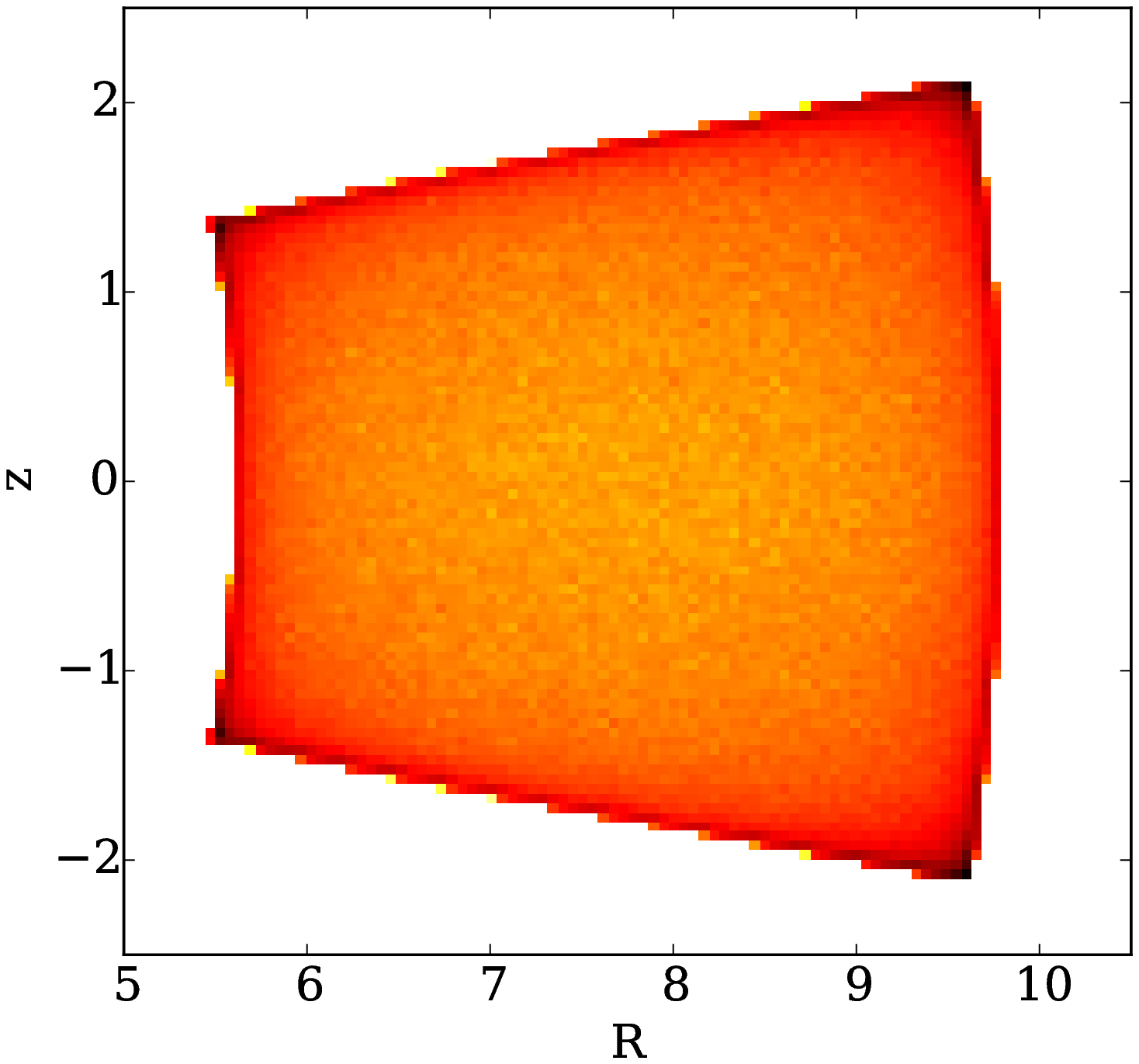}}
}
  \caption{ Plots illustrating the differences when near resonance between orbits found by
    following the expected path on a fitted torus and found by
    Runge-Kutta integration. The upper and lower
    panels show different orbits, both near a 1:1 resonance. The upper
    panels are for a torus with actions
    $\vJ=(0.05,0.085,0.9)\kpc^2\Myr^{-1}$ and frequency ratio
    $\Omega_z/\Omega_r=0.99674$ while the lower panels are for a torus 
    with actions $(0.1,0.07,1.7)\kpc^2\Myr^{-1}$ and frequency ratio $1.0015$. In each
    case the leftmost panel shows  $R$ and $z$ as a function of $t$
    for the torus (black) and for the integrated orbit (red); the
    second-left panel shows the surface of section 
    (i.e. values of $R$ and $v_R$ as the orbit passes $z=0$ with
    $v_z>0$), again with that for the torus in black and that for the
    integrated orbit in red. The density plots show the density in the
    $(R,z)$ plane
    associated with the integrated orbit (second-right) or fitted
    torus (rightmost). Darker colours correspond to higher densities.
\label{fig:res_orbit2}
}
\end{figure*}

Near the Sun the most important resonance from the perspective of trapping is
that for which $\Omega_r=\Omega_z$. At any given energy, satisfaction of this
condition on some orbits is guaranteed by the fact that for nearly
circular orbits $\Omega_z>\Omega_r$ because the disc is massive and thin,
while orbits that make large excursions in $z$, and thus perceive only a
mildly flattened potential, have $\Omega_z<\Omega_r$. 

Each row of Fig.~\ref{fig:res_orbit2} shows an orbit that is trapped by the
resonance $\Omega_r=\Omega_z$.  The orbit shown in the upper row is trapped
such that the $R$ and $z$ oscillations are always roughly in
phase,\footnote{In phase in the sense that smallest $z$ occurs at smallest
$R$. But with \codename's zero-points for $\vtheta$ this implies
$\theta_r-\theta_z\simeq\pi/2$.} so every
time the star moves up through the Galactic plane, it is moving outwards. In
consequence of this, the consequents of this orbit are confined to the upper
half ($v_R>0$) of the surface of section (second panel from the left
of Fig.~\ref{fig:res_orbit2}).

The orbit shown in the lower row of Fig.~\ref{fig:res_orbit2} is trapped
such that its $z$ oscillations lead its $R$ oscillations by $\sim\pi/2$ in
phase, with the consequence that the orbit has a definite sense of
circulation in the $(R,z)$ plane. The white square at the centre of the orbit
is analogous to the unvisited circle at the centre of a familiar rosette
orbit in an axisymmetric potential. In the surface of section (second panel
from the left), the orbit's consequents are confined to the left half of the
diagram because the orbit is always near pericentre when it rises through the
plane. 

On account of the potential's symmetry in $z$, each of the two plotted orbits
is associated with a mirror orbit obtained by mapping $z\to-z$. In the third
column of Fig.~\ref{fig:res_orbit2}, the mirror of the top orbit slopes from
upper left to lower right, and in the second column its consequents would lie
in the lower half of the diagram. The mirror of the lower orbit in
Fig.~\ref{fig:res_orbit2} circulates in the opposite sense, so it would look
the same as that plotted in the third column, but its consequents would be
confined to the right side of the panel in the second column. 

In the PM11 potential, trapping such that the relative phases of the $R$ and
$z$ oscillations librates around 0 or $\pi$ occurs at
$J_\phi\la1.15\kpc^2\Myr^{-1}$ (upper panels of Fig.~\ref{fig:res_orbit2}),
while at larger values of $J_\phi$ the relative phase of the $R$ and $z$
librates around $\pm\pi/2$. Thus a surface of section that showed both orbits
such as those plotted in Fig.~\ref{fig:res_orbit2} and their mirrors would
show a pair of islands that touched at two points on the $R$ axis when
$J_\phi<1.15\kpc^2\Myr^{-1}$ and at top and bottom when $J_\phi$ is larger.
Orbits with $J_\phi=1.15\kpc^2\Myr^{-1}$ are not trapped by the 1:1
resonance.

The black curves in the surfaces of section would run through these islands.
They show the cross section of the torus that has the actions these orbits
would have if they were not resonant.  In reality {\it no\/} orbit has these
values of $J_r$ and $J_z$ -- \cite{JJBSp84} called these ``missing actions''.
Since these resonant orbits are quasiperiodic, they do have actions, but one
of these actions measures the extents of their libration around the trapping
resonant orbit and not the extent of their radial and vertical excursions. 

In summary, with \codename\ we can generate tori for any values of the
actions. Even though the frequencies of some tori will be commensurable, all
tori represent non-resonant orbits in the sense that the tori impose no phase
relationship between the $R$ and $z$ oscillations. For many values of $\vJ$,
non-resonant orbits exist with these actions, and \codename\ should be able
to generate these orbits to arbitrary precision, with the consequence that on
the computed tori the \rms\ fluctuation in $H$ can be made arbitrarily small.
In some ranges of $\vJ$ it is impossible to fit tori of the specified form
into surfaces of constant $H$. Consequently, on any tori with these values of
$\vJ$ that \codename\ constructs, $H$ will fluctuate by a non-negligible
amount. 

If we use \codename\ to foliate phase space with tori, and assign to the
torus $\vJ$ the average value, $\overline{H}(\vJ)$ that $H$ takes on that
torus, then $\overline{H}$ is an integrable Hamiltonian for which we have
angle-action coordinates. The difference 
\[
\Delta(\vtheta,\vJ)\equiv H(\vtheta,\vJ)-\overline{H}(\vJ)
\]
 is a small perturbation. The existence of this perturbation to an integrable
Hamiltonian explains why real orbits are trapped by the resonance, and the
trapping process can be studied with first-order Hamiltonian perturbation
theory \citep{KaJJB94:PhysRev,Ka95:closed}. In the example just discussed,
$\Delta(\vtheta,\vJ)$ at a given value of $\vtheta$ evidently changes sign as
$J_\phi$ varies across the value $J_\phi=1.15\kpc^2\Myr^{-1}$, and
vanishes at this particular value of $J_\phi$.

Crucial for the construction of $\overline{H}$ is that \codename\ creates
tori for adjacent values of the actions which touch but do not cross: if we
are to have a global system of angle-action coordinates, each point in phase
space must be on precisely one torus. This requirement that tori constructed
for neighbouring values of $\vJ$ do not cross is non-trivial.

The existence of resonant trapping has implications for the choice of the
tolerance parameter $\tolJ$. First, given the logical impossibility of
driving $\delta H$ to zero when trapping occurs, very small values of $\tolJ$
are inappropriate: the specified value of $\tolJ$ should be large enough for
the task \codename\ is set to be feasible. Moreover, in parts of action space
that correspond to orbits that are strongly influenced by a resonance that
nevertheless does not trap them, it may be expedient to avoid small values of
$\tolJ$ and closely fitting tori in order to prevent adjacent tori crossing.
That is, if you want a clean foliation of phase space with tori, you should
limit the extent to which resonances distort tori by using a moderate value
of $\tolJ$. We plan to examine more carefully the many issues raised by
orbital trapping in a forthcoming paper.

\subsection{Star-streaming in the Rz plane?}\label{sec:Rzstream}

Given that there are orbits with a well defined sense of circulation
in the $Rz$ plane, one could construct an equilibrium Galaxy model in which
$\langle{v_z}\rangle\ne0$ in the Galactic plane. This is a remarkable possibility.
Naively one would interpret a non-vanishing value of $\langle{v_z}\rangle$ in the
Galactic plane as evidence for a non-equilibrium process, such as  flapping
or warping. Careful examination of the range of values of $|v_z|$ in which
there were more stars  moving up/down than down/up could distinguish between
the equilibrium and non-equilibrium possibilities: in the former case the
up/down asymmetry would be confined to velocities characteristic of the 1:1
resonance. Moreover, a net upward flow of stars at one radius would be
balanced by a net downward flow at a small/larger radius in a velocity range
that could be accurately predicted given knowledge of the potential. The
Sun's motion perpendicular to the plane is measured relative to stars that
are too tightly confined to the plane to be affected by the resonance, so the
conventional value, $v_z=7.25\kms$, is secure.

As a massive star cluster is tidally disrupted and donates its stars to the
thick disc, it will be moving either down or up through the disc at a
particular radius. If this radius is one associated with orbits appropriately
trapped by the 1:1 resonance, its stars will subsequently have a net sense of
circulation in the $Rz$ plane. A burst of star formation induced by a
massive, dense cloud hitting the disc at an appropriate radius could likewise
induce a net circulation in part of the thick disc. Consequently, hunting for
this phenomenon is star catalogues seems a worthwhile activity.

\section{Conclusion} \label{sec:conclude}

We have presented a package of {\tt C++} code, \codename, that allows the
user to find the orbital torus associated with given values of the actions
$\vJ$ in any axisymmetric potential $\Phi$. Once a torus has been fitted, it
is trivial to determine the star's position, velocity and contribution to the
local density for any value $\vtheta$ of the star's angle variables.
Alternatively, one can choose a location $\vx$ and determine whether the star
will ever visit that location, and if so with which velocities and how much it
will contribute to the density there. Since the
orbital frequencies are available, the star's trajectory from any location on
the orbit can be readily recovered without integrating the equations of motion.
Equally, one can recover the curve on which its consequents will lie in a
surface of section.

On a typical single processor $\sim10$ tori can be fitted per second.
Evaluating quantities for a given torus takes just a few ms. Although a
torus completely describes a star's motion for all times, it can be stored in
$\sim100$ numbers. \codename\ provides for the construction of new tori by
interpolation on fitted tori, which is an exceedingly fast operation.

Initial conditions that are in statistical equilibrium in a given
potential are readily chosen with tools provided in \codename: in addition to
torus-finding and manipulating routines, the package includes distribution
functions for realistic discs and a MCMC sampler.

Our discussion of the connection between resonant trapping and torus mapping
has been rather superficial because this topic merits a paper on its own. The
key points are (i) that torus mapping provides accurate representation of all
orbits that are not resonantly trapped, and (ii) that it provides the means
to construct an integrable Hamiltonian that can be used to compute resonant
trapping precisely, and to gain a deeper understanding of this phenomenon and
its implications for galactic dynamics. Examination of trapping by the 1:1
resonance is a realistic Galactic potential indicates that at some radii and
amplitudes of vertical excursions, stars may have a well defined sense of
circulation in the $Rz$ plane.

The \codename\ package can be downloaded from
github.com/PaulMcMillan-Astro/Torus. It contains a number of example
programmes, including ones that produce the data plotted here.  Development
of \codename\ continues, and improvements and additions will be made from
time to time. Relevant software can be downloaded from
github.com/GalacticDynamics-Oxford/ABGal, that includes provision for linking
\codename\ to code implementing the St{\"a}ckel fudge, so that approaches
combining the techniques \citep[see e.g.][]{SaJJB15:Triaxial} are easy to
use.

\section*{Acknowledgements}

Several colleagues have made major contributions to this project.  In the
early 1990s Colin McGill and Mikko Kaasalainen contributed to most of the key
ideas and wrote early \textsc{fortran} codes. In 1995--6 Walter Dehnen
rewrote Kaasalainen's code in \textsc{c++} and most of the class structure of
\codename\ derives from this rewrite. The project has been generously
supported by successive UK research councils: SERC prior to 1994, PPARC from
1994--2007, and latterly by STFC under grants ST/G002479/1, ST/J00149X/1, and
ST/K00106X/1.

The research leading to these results has also received funding from the European Research
Council under the European Union's Seventh Framework Programme (FP7/2007-2013)/ERC
grant agreement no.\ 321067.

PJM gratefully acknowledges financial support from the Swedish National Space
Board and the Royal Physiographic Society in Lund.

We thank Payel Das and Eugene Vasiliev for useful comments on a draft.

\bibliographystyle{mn2e} \bibliography{new_refs}

\appendix
\section{what TM actually does}\label{app:does}

Classically, angle-action coordinates $(\vtheta,\vJ)$ are obtained by solving the
Hamilton-Jacobi equation for the generating function $S(\vx,\vJ)$ of the
canonical transformation $(\vx,\vv)\leftrightarrow(\vtheta,\vJ)$. Since
analytic solutions of the Hamilton-Jacobi equation are not available for
generic potentials, \codename\ obtains the required transformation by
compounding up to three canonical transformations:
 \[
(\vtheta,\vJ){\stackrel{S_\vn}\longrightarrow}(\vthetaT,\vJT){\stackrel{\rm HJ\ 
eqn}\longrightarrow}(\vxT,\vvT){\stackrel{\rm Point\
transf}\longrightarrow}(\vx,\vv).
\]
 The point transformation $\vxT(\vx)$ is employed only if $J_z/J_r>0.05$ and
an attempt to fit a torus without a point transformation has failed. If it is required, \codename\ finds a point
transformation that
maps the relevant circular orbit $\JT_r=0$ in the toy potential $\PhiT$ into
the shell orbit $J_r=0$ in the Galactic potential. Otherwise it just uses the
identity transformation. Next it solves for the
coefficients $S_\vn$ of the generating function (\ref{eq:defsS}) and the
optimum parameters of $\PhiT$. Finally it determines the derivatives $\p
S_\vn/\p\vJ$, which are required to map between toy and true angle variables,
\begin{equation} \label{eq:thinthT} \vtheta = \vthetaT + 2
  \sum_{\vn>0} \frac{\partial S_\vn(\vJ)}{\partial \vJ} \sin{(\vn
    \cdot \vthetaT)} .
\end{equation}

\begin{figure*}
  \centerline{
    \resizebox{0.25\hsize}{!}{\includegraphics{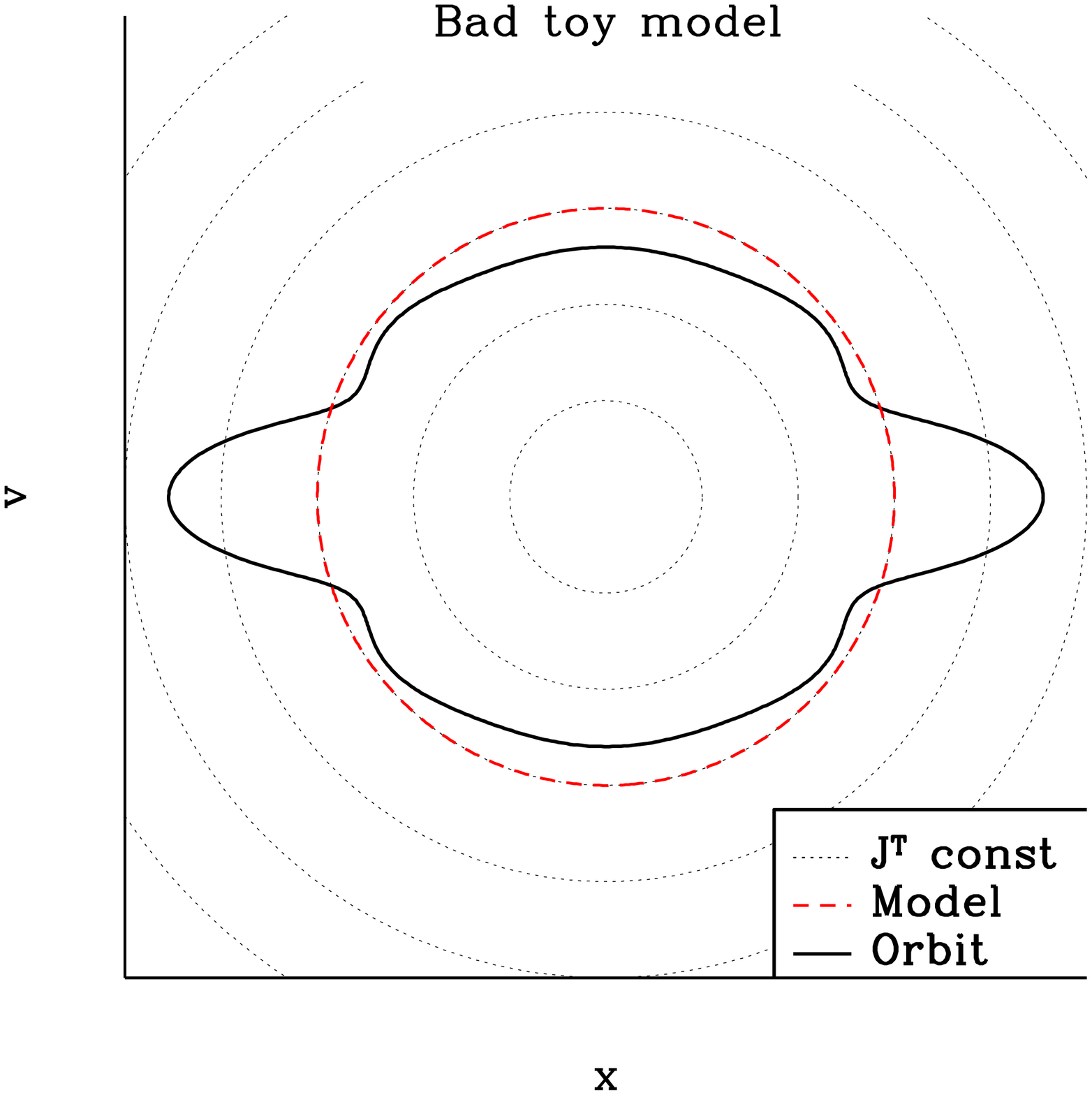}}
    \resizebox{0.25\hsize}{!}{\includegraphics{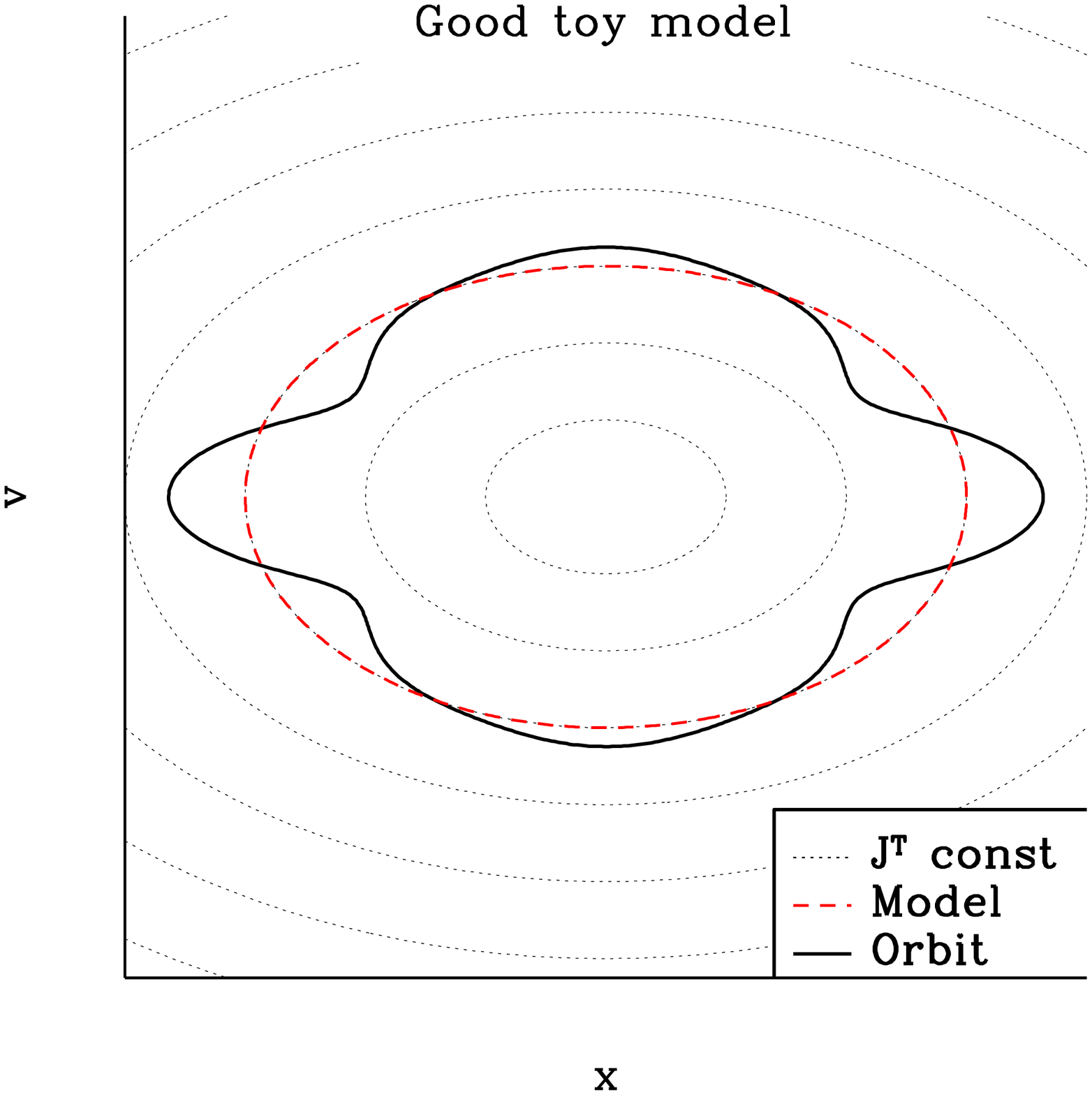}}
    \resizebox{0.25\hsize}{!}{\includegraphics{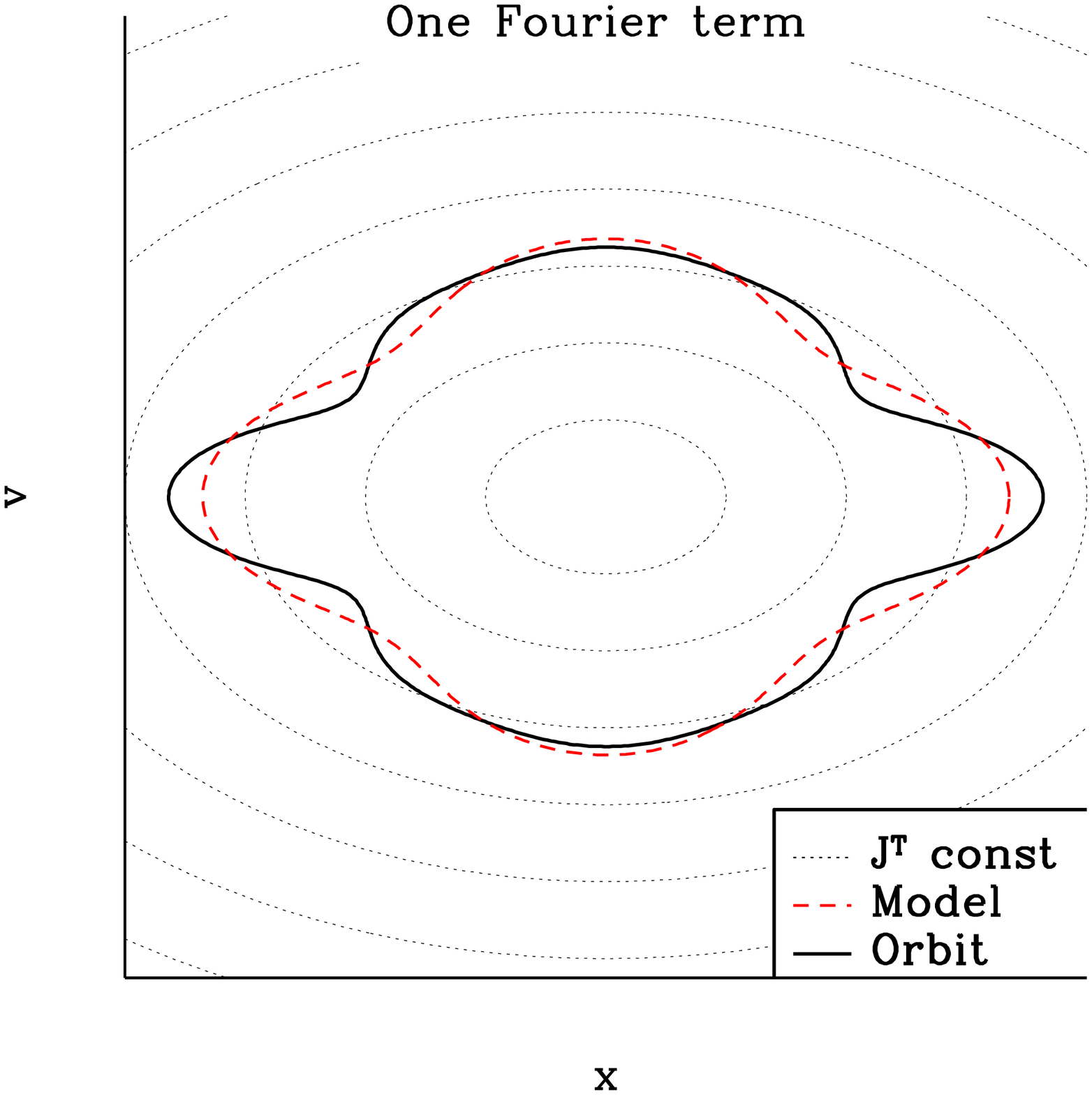}}
    \resizebox{0.25\hsize}{!}{\includegraphics{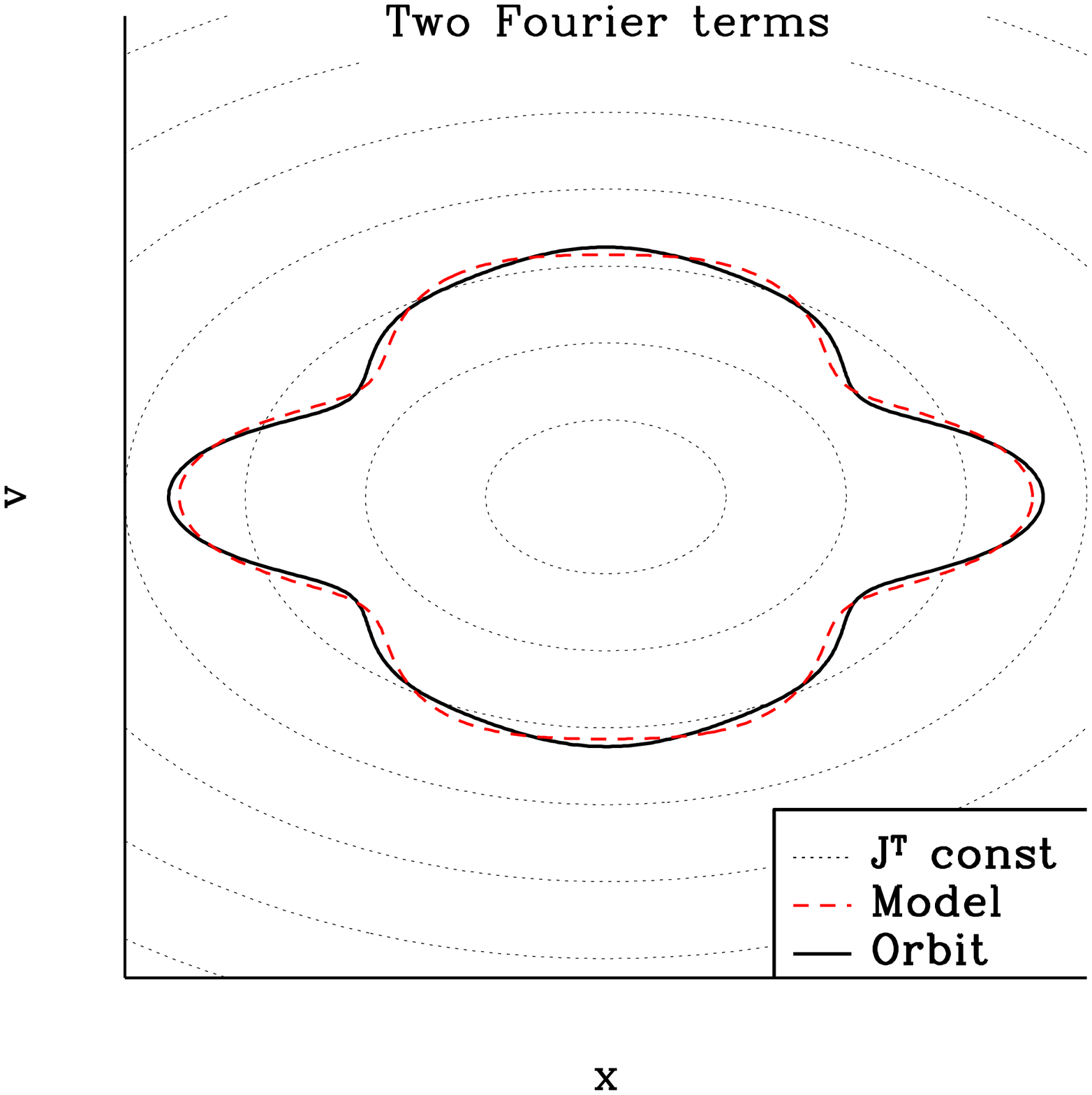}}
}
  \caption{ 
    Illustration of the torus method in the 1D case. Angle-action
    coordinates in the toy potential provide a polar coordinate system
    (dotted lines) in the $q$-$p$ plane
    with the actions giving a radius and the angle as a polar
    angle (left). To fit the true orbit with actions $J'$ (solid line)
    we start with the line $J^T=J'$ in a toy potential (red dashed -
    first panel), then adjust a parameter of the toy potential such
    that this provides a better fit to the orbit (second panel). 
    We then add terms periodic in $\theta^T$ that conserve
    the total area within the red dashed curve (i.e. conserve the
    action), which allow us to describe the true orbit with greater
    and greater accuracy as we add more terms (third and fourth panels). 
\label{fig:scheme_sn}
{\sl Modify (i) to use $(x,v_x)$ instead of (q,p); (ii) delete coord curves from 3rd
\& 4th panels.}}
\end{figure*}

Fig.~\ref{fig:scheme_sn} illustrates the central idea of torus mapping
with a simple one-dimensional example that does not require a point
transformation. In each panel the full black curve
shows the phase space trajectory of the orbit we seek to model. The
angle-action coordinates of a default toy potential $\PhiT$ constitute a system of
polar coordinates $(\thetaT,\JT)$ for phase space. In the leftmost panel this
system is symbolised by a series of circles of constant $\JT$.  The dashed
red curve shows the coordinate curve that encloses the same area as the
orbit's curve, and therefore has $\JT=J$, the orbit's action.

In the next panel the parameters of $\PhiT$ have been adjusted to bring the
dashed red curve $\JT=J$ into closer alignment with the orbit. In the next
panel a generating function with single coefficient $S_\vn$ has been used to
deform the red curve whilst leaving the area it encloses constant. In the
final panel a generating function with several $S_\vn$ has been used to bring
the red curve into alignment with the black curve to the precision required
by the user.

\subsection{The generating function} \label{sec:GF} 

\begin{figure}
  \centerline{\resizebox{\hsize}{!}{\includegraphics{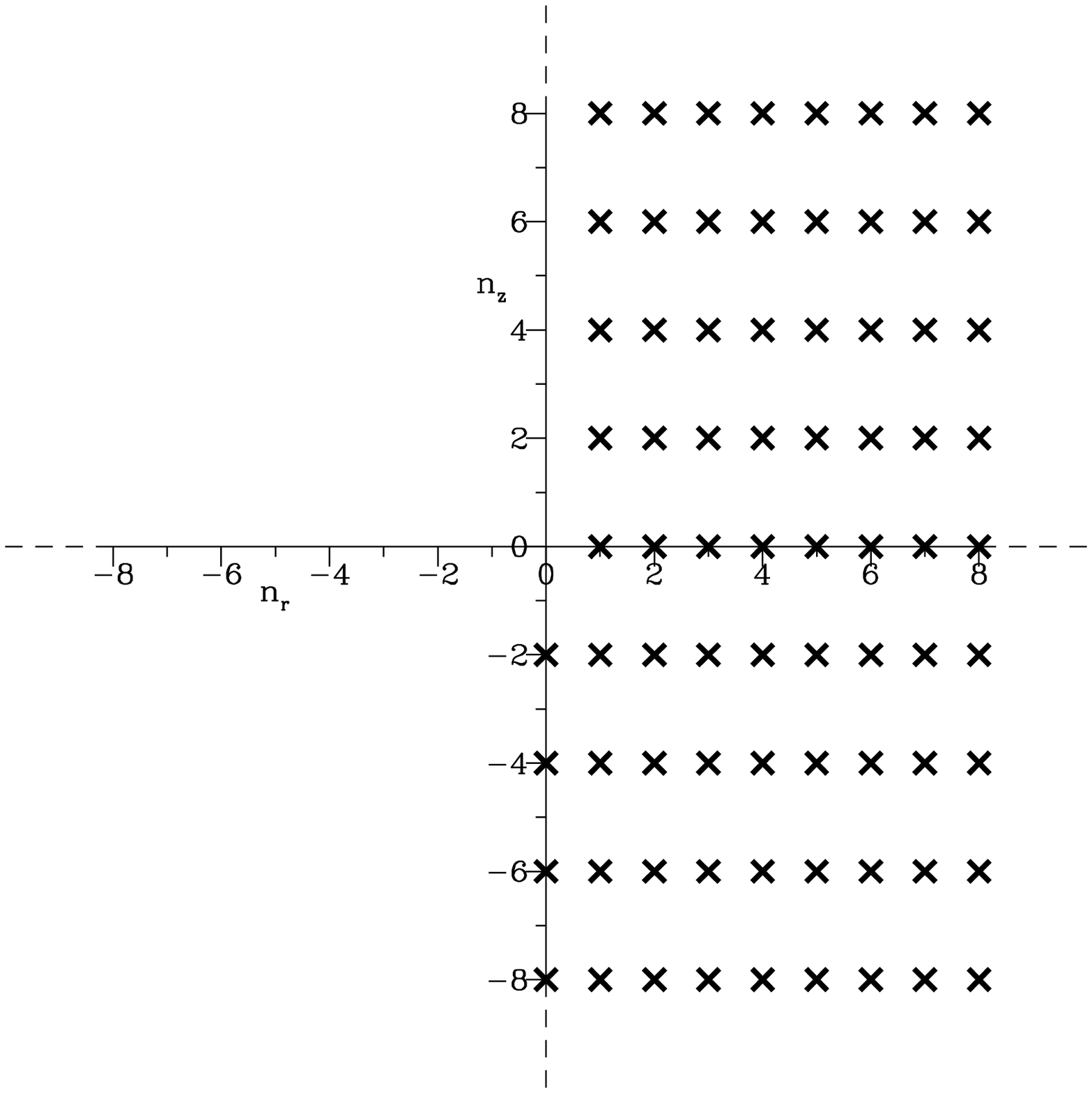}}}
  \caption{ Figure illustrating the pattern of allowed values of
    $(n_r,n_z)$ in the generating function (eq.~\ref{eq:GF}). The
    figure stops at $n_i=8$, but terms can be used (if they follow the
    same rules) for $n\rightarrow\infty$
    \label{fig:allown}
  }
\end{figure}

Given that $\Phi$  is mirror symmetric in the Galactic plane,
the general form (\ref{eq:defsS}) of the generating function can be specialised to
\citep{McGJJB90}
\begin{equation} \label{eq:GF} S(\vthetaT,\vJ) = \vthetaT \cdot \vJ
  + 2 \sum_{\vn} S_\vn(\vJ)\sin{(\vn \cdot \vthetaT)}.
\end{equation}
 where the $S_\vn$ are real, $n_r>0$ for $n_z\ge0$, and only even values of $n_z$ occur --
the allowed values of $\vn$ are illustrated in
Figure~\ref{fig:allown}.
At any point $\vthetaT$ on the torus with actions
$\vJ$, the toy actions are then given by
\begin{equation} \label{eq:JTinJ}
 \vJT = \vJ + 2 \sum_{\vn>0}\vn  S_\vn(\vJ) \cos{(\vn \cdot \vthetaT)}.
\end{equation}

For given $\vJ$ and a regularly-spaced grid of values $\vthetaT$, \codename\
evaluates $\vJT$ and uses the analytic relation between $(\vthetaT,\vJT)$ and
$(\vxT,\vvT)$ and then the point transformation $\vx(\vxT)$ to evaluate
$H(\vx,\vv)$, and thus determines the variance $(\delta H)^2$ of $H$ around
the torus. The Levenberg-Marquardt algorithm \citep{NumRec} is then used to
adjust the $S_\vn$ and possibly the parameters of $\PhiT$ to minimise
$(\delta H)^2$.  The Levenberg-Marquardt algorithm requires the derivatives
of $H$ with respect to the $S_\vn$ and the parameters of $\PhiT$, and these
are all analytically computed using the chain rule. On account of the
symmetries of the potential, the grid in $\vthetaT$ can be restricted to
$0\leq\thetaT_r<\pi$ and $0\leq\thetaT_z<\pi$.

\codename\ starts with a small number of coefficients $S_\vn$, by default
those with $\vn=(1,0)$, $(2,0)$ $(3,0)$, $(0,-2)$, $(0,-4)$, $(1,2)$,
$(1,-2)$, $(1,4)$.  After an optimisation of these coefficients it adds
coefficients at values of $\vn$ that are adjacent to points at which
$|S_\vn|$ has been found to be above a threshold value that is a fixed
fraction of the largest $|S_\vn|$.  Each $S_\vn$ is set to zero on its
introduction.  When additional $S_\vn$ are introduced, it may be necessary to
make the grid in $\vthetaT$ denser to ensure that the sampling density around
the torus comfortably exceeds the Nyquist frequency associated with the
largest value of $|\vn|$. Consequently, the computational cost of each
optimisation step increases quite rapidly as the number of $S_\vn$ increases.
The number of optimisation steps is limited, by default to 10. After 10 steps
there are typically $\sim700$ non-zero $S_\vn$.

Once a satisfactory variance in $H$ has been achieved, \codename\ solves for
the derivatives of the $S_\vn$, which appear in the relation
(\ref{eq:thinthT}) between the toy and true angles.  The derivatives are
obtained analytically rather than by finite differences.  Currently
\codename\ does this using the algorithm introduced by
\cite{KaJJB94:MNRAS}.\footnote{Angle fitting will shortly be shortly changed
(Vasiliev et al., in preparation) to the algorithm introduced by
\cite{JJBKu93} as modified by \cite{LaaksoKaas}.} From several initial
conditions drawn from the torus, \codename\ integrates the equations of
motion for $M$ time-steps and computes the toy angles after each time-step.
These must satisfy
\begin{equation}
  \vtheta(0) + \vOmega t_i  = \vthetaT(t_i) + 2\sum_{\vn>0}
  \frac{\partial S_\vn(\vJ)}{\partial \vJ} \sin[\vn\cdot\vthetaT(t_i)]
\end{equation}
 for $i=1,\ldots,M$. Hence each integration yields $3M$ equations in which
$\vtheta(0)$, $\vOmega$, and ${\p S_\vn(\vJ)}/{\p\vJ}$ appear as unknowns.
The equations are linear in the unknowns and for $M\gg1$ the number of
available equations increases much faster than the number of unknowns. We
solve these equations in a least-squares sense.  We refer to this process as
an ``angle fit''.

The various steps just described are each implemented by an instance of a
distinct class:

\begin{itemize}

\item
The $\vn$ and corresponding
$S_\vn$, are stored and manipulated by an instance {\tt Sn} of the class {\tt GenPar}.

\item 
The $\vn$ and the corresponding values $\partial S_\vn/\p\vJ$ are
stored and manipulated by an instance {\tt A} of the class {\tt AngPar}.

\item
An instance {\tt GF} of the  class {\tt GenFnc} handles the mapping from $(\vJ,\vthetaT)$ to
$(\vJT,\vthetaT)$ for a fixed value of $\vJ$. It also finds the derivatives
$(\p\vJT/\p\vthetaT)_\vJ$. {\tt Sn}  is a member of {\tt GF}.

\item 
An instance {\tt GFF} of the class {\tt GenFncFit} performs the same tasks as
{\tt GF} but focuses on the transformations required when performing an
action fit, where we have a known, regular grid
of points in $\vthetaT$ at which the transformation has to be calculated many
times. {\tt GFF} also provides the derivatives $\p\vJT/\p S_\vn$.

\item
An instance {\tt AM} of the class {\tt AngMap} handles the mapping between
$(\vJ,\vtheta)$ and $(\vJ,\vthetaT)$ (eq.~\ref{eq:thinthT}). This mapping
works in either direction, and the values
$\partial\theta_i/\partial\theta^T_j$ can also be found. {\tt A} is a member
of {\tt AM}.

\item
An instance {\tt T} of the class {\tt Torus} has {\tt GF} and {\tt AM}
as members.

\end{itemize}

\subsection{The Toy Potential} \label{sec:ToyIso}

The job of the toy potential is to provide analytic angle-action coordinates
$(\vthetaT,\vJT)$.
Candidates include the harmonic oscillator potential, H\'enon's isochrone
\citep{McGJJB90}, and a St\"ackel potential \citep{LaaksoKaas}. \codename\
uses the generalised effective isochrone potential
\begin{equation} \label{eq:isochrone}
  \Phi_{\mathrm{eff}}^T(r,\vartheta) =\frac{-GM}{b+\sqrt{b^2 +
      (r-r_0)^2}} +
  \frac{L_z^2}{2\left[(r-r_0)\sin\vartheta\right]^2},
\end{equation}
where $\vartheta$ is the usual spherical polar coordinate (not to be confused
with a dynamical angle coordinates), and $M$, $b$, $L_z$ and $r_0$ are the
potential's parameters. Since we are modelling the motion in the $(R,z)$
plane, we can treat $L_z$ as a parameter of the toy effective potential that
is distinct from the actual $z$-component of angular momentum $J_\phi$.
Appendix A2 of \cite{McGJJB90} gives the formulae needed to compute
$(R,z,v_R,v_z)$ from $(\theta_r,\theta_z,J_r,J_z)$ with the provisos: (i) for
$\cos\theta$ in McGill \& Binney read $\sin\vartheta$, (ii) for $J_\vartheta$
read $J_z$, and (iii) for $r$ read $r-r_0$.  Expressions for $\thetaT_\phi$
are given below. If $\thetaT_\phi$ is to increase in the same sense as
$\theta_\phi$, $L_z$ and $J_\phi$ must have the same sign.

The parameters of $\PhiT$ actually used by \codename\
are $ L_z, r_0$,
\[
\gamma\equiv\sqrt{GM}\hbox{ and }\beta\equiv\sqrt{b}.
\]
This parametrisation ensures that  unphysical negative values of $M$
and $b$ cannot be inadvertently chosen during torus fitting.

The class {\tt ToyMap} is a base class for any mapping between
$(\vthetaT,\vJT)$ and $(\vxT,\vvT)$. Having this base class facilitates
future use  of the tori of a three-dimensional harmonic oscillator, but
currently the only class of this kind that is implemented is {\tt
ToyIsochrone}, which specifies the mapping in case of the generalised
effective isochrone potential. It also specifies the calculation of the
partial derivatives of $(\vxT,\vvT)$ with respect to $\vJT$, $\vthetaT$ and
the parameters of the toy potential.

\subsection{Point transformation} \label{sec:PT} 

Unless $J_r/J_z$ is small, a torus of $\PhiT$ can be mapped onto a
torus of the Galactic potential using only the generating function
(\ref{eq:GF}). Unfortunately for small values of $J_r/J_z$ a more
sophisticated treatment is necessary.

\begin{figure}
  \centerline{\includegraphics[width=.4\hsize]{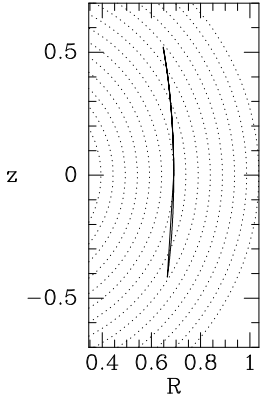}}
  \caption{
    A target shell orbit in the $(R,z)$ plane ({solid} line) and shell
    orbits in the toy potential (which are all on circles centred on the
    origin, {dotted} lines). The generating function of
    Section~\ref{sec:GF} is incapable of mapping the toy orbits on to
    the target orbit. 
\label{fig:shellprobfig}
}
\end{figure}

As $J_r\rightarrow0$ eq.~(\ref{eq:JTinJ}) implies negative values of $\JT_r$
unless all the $S_\vn\to0$, so the transformation generated by
$S(\vthetaT,\vJ)$ tends to the identity. However, while the shell orbits
$\JT_r=0$ of the isochrone potential lie on spheres, the shell orbits $J_r=0$
of a flattened Galactic potential do not (Figure~\ref{fig:shellprobfig}).
Consequently, along a shell orbit in a Galactic potential $r$ oscillates, so
$p_r\neq0$. But if $p_r\neq0$, then $\JT_r\neq0$ also, because $p_r$ is
non-zero only on eccentric orbits. It follows that the generating
function (\ref{eq:GF}) is insufficiently general to
yield image tori that have small radial actions.

\cite{KaJJB94:MNRAS} solved this problem by
combining the canonical transformation generated by eqn.~(\ref{eq:GF}) with a point
transformation $\vx\to\vxT(\vx)$ of the form
\[
(r,\vartheta)\rightarrow(\rT,\varthetaT)=
\left(\xi(\vartheta)r,\;\eta(r)\zeta(\vartheta)\right).
\]
Here $\xi$, $\eta$ and $\zeta$ are functions to be determined such that the
shell orbit of $\PhiT$ with the given $(J_z,J_\phi)$ is mapped into the
corresponding shell orbit of the Galactic potential. Qualitatively, we want
$\eta\sim1$ and $\zeta\sim\vartheta$; $\xi$ is chosen such that the shell
orbit of the target potential is $\rT=a$, where $a$ is the radius of the toy
shell orbit.  Since the radius and angular extent of the shell orbit in
$\PhiT$ depends on the parameters of $\PhiT$, the latter must be fixed before
the point transformation is determined, and if it changed by the
Levenberg-Marquardt routine as it minimises $\delta H$, the shell orbits will
no longer be mapped into each other and tori $J_r=0$ will cease to be
accessible.  We set $r_0=0$,
$L_z=J_\phi$, and (rather arbitrarily) $\beta=\sqrt{3}$. $\gamma$ is chosen
such that in $\PhiT$ the orbit has $r=1$ and in the formulae below $a$ can be
set to unity. 

\begin{figure}
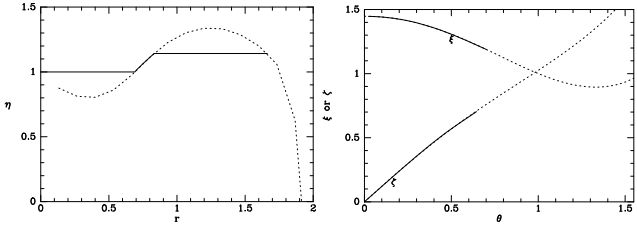

  \centerline{\includegraphics[width=.49\hsize]{plots/yofr.ps}
    \includegraphics[width=.49\hsize]{plots/xoftheta.ps}}
  \caption{
    The functions $\eta(r)$
    (left) and $\xi(\theta)$, $\zeta(\theta)$ (right) that define the new
    $(r,\theta)$ coordinates. The full curves show results obtained by
    integrating the defining equations and then adding extra points to fill out
    the range, while the dashed curves show Chebychev-polynomial $L^2$ fits to
    the points defined by these curves.
\label{fig:xietazeta}
}
\end{figure}

Numerical integration of the Galactic shell orbit yields the radius of
this orbit, $\rs(\vartheta)$ and we immediately have
\[
\xi(\vartheta)\equiv {a\over\rs(\vartheta)}
\]
because then on the orbit $\rT=\xi\rs=a$ as
required. The upper curve in the right-hand panel of Fig.~\ref{fig:xietazeta} shows $\xi(\vartheta)$ for the
shell orbit plotted in Figure \ref{fig:shellprobfig}.

The generating function of our point transformation is
\[\label{eq:basicGFS}
S(r,\vartheta,\pT_r,\pT_\vartheta)=\xi r\pT_r+\eta\zeta \pT_\vartheta,
\]
so
\[ \label{eq:PTmomenta} p_r=\xi\pT_r+{\rd\eta\over\rd r}\zeta
\pT_\vartheta\quad;\quad
p_\vartheta={\rd\xi\over\rd\vartheta}r\pT_r+\eta{\rd\zeta\over\rd\vartheta}\pT_\vartheta.
\]
From our integration of the target shell orbit we know
$p_r(\vartheta)$ and $p_\vartheta(\vartheta)$ along this orbit. On the toy
potential's shell
orbit $p_r^T=0$ and
\[
\pT_\vartheta(\varthetaT)=\sqrt{L^2-{L_z^2\over\sin^2\varthetaT}}.
\]
Finally we use the known relation $\rs(\vartheta)$ on the target shell
orbit to treat $\eta$ as a function of $\vartheta$. Then equations
(\ref{eq:PTmomenta}) yield coupled o.d.e.s for $\eta$ and $\zeta$
\[ {\rd\eta\over\rd\vartheta}={p_r\over \zeta \pT_\vartheta}
{\rd\rs\over\rd\vartheta}\quad;\quad
{\rd\zeta\over\rd\vartheta}={p_\vartheta\over \eta \pT_\vartheta}.
\]
Figure~\ref{fig:xietazeta} shows the functions $\eta$ and $\zeta$ that
are obtained by integrating these equations. \footnote{In practice
  \codename\ changes these o.d.e.s to ones with a new independent
  variable $\psi$ before integrating the equations, where
  $\vartheta=\vartheta_{\rm max}\sin\psi$, with $\vartheta_{\rm max}$
  the largest value of $\vartheta$ attained on the closed orbit.}

So far we have obtained the functions $\xi$, $\eta$ and $\zeta$ only
in the ranges of $r$ and $\vartheta$ covered by the closed orbit. The
functions are extended beyond this range by adding single points at
the ends of the required ranges, and then fitting a sum of Chebychev
polynomials to both these points and the points returned by the
numerical integrations. The fitting algorithm does not require the
polynomial to pass through the given points. Rather a function is
constructed from the points by fitting a quadratic to each set of
three points. Then the Chebychev series is fitted in a least squares
sense to this function -- the coefficients of the Chebychev
polynomials are found from integrals of the products of the relevant
Chebychev polynomial and the quadratics and the weight function. These
integrals are done analytically.  By relieving the polynomial of the
obligation to pass through the data points, this techniques avoids the
danger that the polynomial makes wild excursions between data points.

\begin{figure}
  \centerline{\includegraphics[width=.4\hsize]{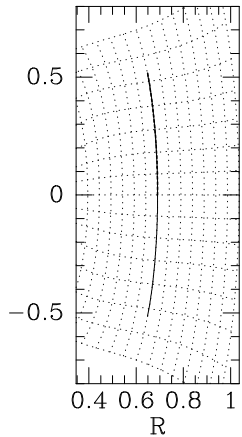}}
  \caption{
    The dotted lines show the
    $(r,\vartheta)$ coordinate system constructed such that a shell orbit in
    the target potential coincides with a curve of constant $r$.
\label{fig:finalcoords}
}
\end{figure}

\begin{figure}
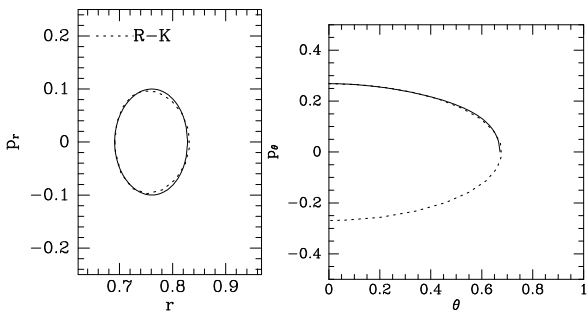

  \centerline{\includegraphics[width=.4\hsize]{plots/rpr.ps}
    \includegraphics[width=.5\hsize]{plots/thetaptheta.ps}}
  \caption{
    Fits in the two position-momentum planes: dotted
    curves numerical integration, full curves images of the toy closed orbit
    under the point transformation.
\label{fig:pmapfig}
}
\end{figure}

Figure~\ref{fig:finalcoords} shows the new coordinates, while
Figure~\ref{fig:pmapfig} shows the fits obtained to the target shell orbit in
the $(r,p_r)$ and $(r,p_\theta)$ planes. The parameters of the point
transform are the coefficients of the Chebychev polynomial.

The class {\tt PoiTra} is the base class for any point transformation
$(\vxT,\vvT)\leftrightarrow(\vx,\vv)$. It uses cylindrical polar coordinates.
If no point transformation is required, \codename\ uses an instance of the
derived class {\tt PoiNone}. If $J_r/J_z$ is small so a point transformation is required,
\codename\ creates an instance of the derived class {\tt PoiClosedOrbit}.

\section{Evaluating $\theta_\phi^T$ in the isochrone
potential}\label{app:isochrone}

Evaluating $\theta_\phi^T$ in the generalised effective isochrone potential
potential (eq. \ref{eq:isochrone}) is not, perhaps, as straightforward
as might appear from the standard texts. We therefore give the details
here.

\subsection{Non-planar orbits}

First let us consider the usual case where $J_\vartheta \neq 0$ (with
$\vartheta$ the usual polar angle).  We have found the values of
$\theta^T_r,\theta^T_\vartheta$ and $\bolJ$ (or equivalently
$r,\vartheta,p_r,p_\vartheta$) in the generalised effective isochrone
potential potential (eq. \ref{eq:isochrone}). For the standard
isochrone potential ($r_0=0$, no term in $L_z$), we have from
equations (3.229) and (3.232) of \cite{GDII} that
\[
\theta_\phi=\phi-u+\hbox{sgn}(L_z)\theta_\vartheta
\]
with $\phi$ the usual azimuthal angle and 
\[ \label{eq:u}
\sin u\equiv\cot i\cot\vartheta,
\] 
with $i$ the inclination given by $i=\arccos(L_z/L)$.

Care needs to be taken when finding $i$ and $u$. $i$ is found in
the toy Hamiltonian, so $L_z$ is the value used as a parameter in
eqn.~(\ref{eq:isochrone}), and $L = J_\vartheta^T + |L_z|$. 

Equation~(\ref{eq:u}) only provides the value of $\sin u$, while  $u$ takes
values throughout the range $(0,2\pi)$, so this is insufficient information. We need to
take note of the fact that $u = \phi-\Omega$, where in this case $\Omega$ is
the longitude of the ascending node, i.e., the line on which an orbit crosses the
$(x,y)$ plane with $\dot{\vartheta}<0$, (again, in the spherical toy potential,
in which the orbital plane does not precess). Assume that $L_z>0$ and follow
the star up from the ascending node (where $u=0$) until $\vartheta$ stops
decreasing. Then it is geometrically clear that $\vartheta=\pi/2-i$, so $\sin
u=1$ and $u=\pi/2$. Now $\vartheta$ starts to increase and in order to ensure
that $u$ continues to increase we need to shift to the other solution of
eqn.~(\ref{eq:u}), $u=\pi-\arcsin(\cot i\cot\vartheta)$. 

In summary for $L_z>0$ we take
\[\label{eq:u_full}
 u =\begin{cases}
   \arcsin(\cot i\cot\vartheta)& {\rm for\ } \dot\vartheta<0, \\
   \pi-\arcsin(\cot i\cot\vartheta) & {\rm otherwise.}\end{cases}
\]
With this definition, $-\pi/2 < u < 3\pi/2$.

When $L_z<0$, $i>\pi/2$ and $u$ decreases from zero as the star rises
through the ascending node, just as $\phi$ decreases, and eq~(\ref{eq:u}) still
applies.

\subsection{Planar orbits}
When $J_\vartheta = 0$, the value of $\theta_\vartheta$ is undefined,
so eq.~\ref{eq:u_full} is essentially meaningless.  We have to return
to the full expression for the generating function, eqn.~(3.231) of
\cite{GDII}, fixing $J_2 = |J_1| \equiv |L_z|$, and $\vartheta=\pi/2$:
\[
S = \phi L_z  +
\int^r_{r_{\rm min}}\hbox{d}r\,\epsilon_r\sqrt{2H(L_z,J_r)-2\Phi(r)-\frac{L_z^2}{r^2}},
\]
and find $\theta_\phi = \partial S/\partial L_z$. 
Happily, this is very similar to the equations that have already been 
solved to find $\theta_\vartheta$ in the usual case, so we know that we can solve
for $\theta_\phi$:
\begin{eqnarray}
\theta_\phi & = & \phi + \frac{\Omega_\phi}{\Omega_r}\theta_r - 
\hbox{sgn}(L_z)\,\left[\tan^{-1}{\left(\sqrt{\frac{1+e}{1-e}}\tan{(\frac{1}{2}\eta)}\right)}\right.\nonumber\\
& & +\left.\zeta \tan^{-1}\left(
\sqrt{\frac{1+e+2b/c}{1-e+2b/c}}\tan{(\frac{1}{2}\eta)}\right)\right],
\end{eqnarray}
where $e$, $\eta$ and $c$ are as defined in eqn.~(3.240) of \cite{GDII} and 
\[
\zeta = \frac{1}{\sqrt{1+4GMb/L_z^2}}.
\]

\section{Gravitational potentials}\label{app:pot}

\subsection{Multi and predefined potentials}

The class {\tt MultiPotential}  allows one to build a new potential by adding
several previously defined potentials. For example, one can construct the
potential of a disc galaxy by adding the potentials of a bulge, a disc and a
dark halo. 

To create a potential that is the sum of  a
{\tt MiyamotoNagaiPotential} and a {\tt LogPotential}, one writes
\begin{verbatim}
Potential **PotList  = new Potential*[2];
PotList[0] =  new MiyamotoNagaiPotential(M,a,b);
PotList[1] =  new LogPotential(V0,q,Rc);
Potential *Phi = new MultiPotential(PotList,2);
\end{verbatim}

\subsection{User defined potentials}

To create a new potential class, one has to write two new files (the header
file and the main code). As an example, we give below the header and code
files for the Kepler potential (which is identical to {\tt
IsochronePotential} with b=0).

\begin{verbatim}
// file KeplerPotential.h
#include <cmath>
#include "Potential.h"

class KeplerPotential : public Potential {
  double GM;
public:
  KeplerPotential     (const double);
  ~KeplerPotential    () {;}
  double operator()   (const double, 
                        const double) const;
  double operator()   (const double, 
                        double&, double&) const;
  Frequencies KapNuOm (const double) const;
};

// file KeplerPotential.cc
KeplerPotential::KeplerPotential(const double M) {
  GM = Units::G * M;
}
double KeplerPotential::operator(const double R,
                         const double z) const {
  return -GM / sqrt(R*R+z*z);
}
double KeplerPotential::operator(const double R,
                             const double z,
                             double &dPdR,
                             double &dPdz) const {
  double ir2 = 1./(R*R+z*z), ir = sqrt(r2);
  double dPdr = GM * ir2;
  dPdR = dPdr * R*ir;
  dPdz = dPdr * z*ir;
  return -GM * ir;
}

Frequencies KeplerPotential::KapNuOm(const double R) {
  Frequencies epicycle;
  double allfreqs = sqrt(GM/powf(R,1.5));
  // In Kepler potential, all epicycle freqs equal
  epicycle[0] = allfreqs;
  epicycle[1] = allfreqs;
  epicycle[2] = allfreqs;
  return epicycle;
}

\end{verbatim}

\section{User defined distribution functions}
To create a new class of \df\ one adds code to  the file {\tt DF.h}. The
following  example defines a
\df\ that is constant for $J_r<1$, $J_z<1$ and $0<J_\phi<J_{\phi,{\rm
max}}$, and zero elsewhere.

{\obeylines\tt\parindent=10pt
  class Simple\_DF : public DF\{
       public:
\quad       double JpMax;
\quad       int setup(istream\&);
\quad       int setup\_full(istream\&);
\quad       int    NumberofParameters()\{return 1;\}
\quad       void   Parameters(double*);
\quad       double df(Potential*,Actions);
 \};

  inline int Simple\_DF::setup(istream \&ifile)\{  
\quad    char type1; ifile >> type1;
\quad    if(type1!='S') \{
\quad\quad      cerr << "improper input file\textbackslash n";
\quad\quad      return 0;
\quad    \}
\quad ifile >> JpMax;
\quad    return 1;
  \}

  inline void Simple\_DF::Parameters
\qquad\qquad(double* output)\{
\quad    output[0] = JpMax;
  \}

  inline double Simple\_DF::df
\qquad\qquad(Potential* Phi,Actions J)\{
\quad    if(J[0]>1 || J[1]>1) return 0;
\quad    if(J[2]<0 || J[2]>JpMax) return 0;
\quad    return 1/JpMax;
  \} 
}
 \noindent The line 
{\obeylines\tt\parindent=10pt 
DF *sdf=set\_DF(ifile);
}
\noindent will initialise an instance
{\tt sdf} of {\tt Simple\_DF} with {\tt Jpmax = 2} provided (i) the contents
of {\tt ifile} are
 {\obeylines\tt\parindent=10pt
  S
  2
}
 \noindent and (ii) the following code has been added to the
list of possibilities in the definition of {\tt DF::set\_DF} (which is given
at the end of the file {\tt DF.h}):
{\obeylines\tt\parindent=10pt
\quad  if(type1=='S')\{  
\quad\quad    tmpdf = new Simple\_DF;
\quad\quad    tmpdf->setup(ifile);
\quad\quad    return tmpdf;
\quad\}
}

\end{document}